\documentclass[journal]{IEEEtran}

\usepackage{multirow}
\usepackage{multicol}
\usepackage{booktabs}
\usepackage{float}
\usepackage{placeins}
\usepackage{graphicx}
\usepackage{url}
\usepackage{caption}
\usepackage{mathtools}
\usepackage{subcaption}
\usepackage{pbox}
\usepackage{cite}
\usepackage{lipsum}
\usepackage{amsmath}
\usepackage{scalerel}
\usepackage{tikz}
\usetikzlibrary{svg.path}

\definecolor{orcidlogocol}{HTML}{A6CE39}
\tikzset{
    orcidlogo/.pic={
        \fill[orcidlogocol] svg{M256,128c0,70.7-57.3,128-128,128C57.3,256,0,198.7,0,128C0,57.3,57.3,0,128,0C198.7,0,256,57.3,256,128z};
        \fill[white] svg{M86.3,186.2H70.9V79.1h15.4v48.4V186.2z}
        svg{M108.9,79.1h41.6c39.6,0,57,28.3,57,53.6c0,27.5-21.5,53.6-56.8,53.6h-41.8V79.1z M124.3,172.4h24.5c34.9,0,42.9-26.5,42.9-39.7c0-21.5-13.7-39.7-43.7-39.7h-23.7V172.4z}
        svg{M88.7,56.8c0,5.5-4.5,10.1-10.1,10.1c-5.6,0-10.1-4.6-10.1-10.1c0-5.6,4.5-10.1,10.1-10.1C84.2,46.7,88.7,51.3,88.7,56.8z};
    }
}

\newcommand\orcidicon[1]{\href{https://orcid.org/#1}{\mbox{\scalerel*{
                \begin{tikzpicture}[yscale=-1,transform shape]
                \pic{orcidlogo};
                \end{tikzpicture}
            }{|}}}}

\usepackage[hidelinks]{hyperref}

\ifCLASSINFOpdf
\else
\fi

\newcommand*\rot{\rotatebox{90}}

\begin{document}
\title{Fast Generation of Custom Floating-Point Spatial Filters on FPGAs}

\author{Nelson~Campos$^{\textsuperscript{\orcidicon{0000-0001-9709-2703}}}$,~\IEEEmembership{ARM~Ltd,}
        Eran~Edirisinghe$^{\textsuperscript{\orcidicon{0000-0002-7570-3670}}}$,~\IEEEmembership{Keele~University,}
        \\Slava~Chesnokov,~\IEEEmembership{Imaging-CV Ltd},
        and~Daniel~Larkin,~\IEEEmembership{ARM Ltd}
        
\thanks{Nelson Campos is with ARM Ltd.\protect}
\thanks{Slava Chesnokov is with Imaging-CV Ltd.}
\thanks{Daniel Larkin is with ARM Ltd.}%
\thanks{Professor Eran Edirisinghe is with Keele University.}}

\maketitle

\begin{abstract}
Convolutional Neural Networks (CNNs) have been utilised in many image and video processing applications. The convolution operator, also known as a spatial filter, is usually a linear operation, but this linearity compromises essential features and details inherent in the non-linearity present in many applications. However, due to its slow processing, the use of a nonlinear spatial filter is a significant bottleneck in many software applications. Further, due to their complexity, they are difficult to accelerate in FPGA or VLSI architectures. This paper presents novel FPGA implementations of linear and nonlinear spatial filters. More specifically, the arithmetic computations are carried out in custom floating-point, enabling a tradeoff of precision and hardware compactness, reducing algorithm development time. Further, we show that it is possible to process video at a resolution of 1080p with a frame rate of 60 frames per second, using a low-cost FPGA board. Finally, we show that using a domain-specific language will allow the rapid prototyping of image processing algorithms in custom floating-point arithmetic, allowing non-experts to quickly develop real-time video processing applications.
\end{abstract}


%
\IEEEpeerreviewmaketitle

\section{Introduction}
\IEEEPARstart{I}{mage} and video processing applications are complex tasks that require expensive computation. The inherent parallelism of FPGAs enables them to accelerate digital signal processing at the cost of the increased complexity of algorithms due to the additional effort to map the applications to hardware designs \cite{tsigkanos2020high}.

With the increased popularity of convolutional neural networks (CNN), the hardware acceleration of spatial convolution is a relevant topic. The 2D convolutions are usually the bottleneck of CNN architectures \cite{ma2019performance}. GPUs typically perform millions or billions of floating-point computations per second to process image and video data. Compared to GPUs, FPGAs also have the advantage of low power consumption at the cost of an increased time of project development \cite{cong2018understanding}. However, the low level of abstraction makes FPGA implementations more complex than software acceleration using GPUs. One solution to reduce this development time, high-level synthesis (HLS), is an increasing trend to accelerate hardware design. The application, written in a level of abstraction close to GPU implementations, is converted to hardware description using automated tools. While HLS generates accurate functional architectures, often they are not mapped in the most efficient way to save area and power resources \cite{fingeroff2010high}. For this reason, the traditional hardware flow using hardware description languages such as Verilog and VHDL is still resistant in academia and industry. It requires knowledge at a register transfer level to map arithmetic operations with elementary logic gates and flip-flops.

The convolution operation is complex in terms of implementation and leads to massive resource usage. Especially in high-resolution video processing, the amount of memory required to store a frame increases proportionally to the video resolution \cite{campos2017metodologia}\cite{campos2017framework}\cite{monteiro2017energy}. Traditional software implementations store the frame in a bi-dimensional structure using matrix structures. Hardware implementations also map this structure at the cost of external memories due to the high memory storage. A common way to perform the hardware acceleration of convolutions is by using a circular-fashion stream \cite{mohanty2012memory}. The pixel streams are stored using line buffers updated using synchronous counters, leading to significant memory and resource usage savings. Efficient implementations perform a convolution with a kernel of dimensions $N\times N$ using $N-1$ line buffers. In terms of semantics, a line buffer is just a one-dimensional array that reads and writes a pixel within a video line at every clock cycle. The array can be mapped into different memory structures. Flip-flops are used to model memories with multiple read and write in the same clock, but this solution is inefficient in power consumption and silicon. Because of their cheap cost, dual-port RAMs are often choices to model video streams at the expense of having a limited number of reads and writes, requiring additional efforts in the hardware implementation to model the circular buffers \cite{holzer2012optimized}.

\IEEEpubidadjcol

In addition to the mechanics of manipulating a pixel neighbourhood in a raster stream, a convolutional block must perform arithmetic operations within the pixels inside a region. A linear convolution is a pixel-wise operation that usually multiplies each pixel inside a window region by coefficients stored inside a kernel. If the kernel coefficients are constant, then the multiplications can be replaced by shift register operations. However, if the kernel coefficients are not constant values, then the computation will require DSP blocks to perform the product of each pixel by each coefficient. Moreover, adder trees are customised to calculate the additions of the products, targetting optimisation of the data-path arithmetic with the minimum latency and number of operators \cite{ioannou2020high}. Non-linear operations, such as median cross filters, require sorting pixels inside a neighbourhood to reduce noise in a given region \cite{kent2020design}. Hardware implementations of sorting algorithms usually are accomplished using sorting networks. For example, the Bose-Nelson and Batcher's  Merging Network are popular algorithms implemented using swap and compare operations. 

Different strategies have been adopted to meet the project specification of video processing applications with non-linear characteristics. DSP processors based solutions offer a moderate level of complexity in terms of implementation, usually written in C language and with a high level of programmability at the cost of more power consumption and less computational power when compared to ASIC solutions \cite{saponara2005cost}. On the other hand, ASIC implementations provide much more power computation and efficient power consumption. Still, they have high implementation time, usually written in Verilog or VHDL, leading to lengthy development cycles. While ASIC implementations compromise the programmability of the solutions, ASIP and FPGAs are alternative solutions that still maintain the architecture characteristics of ASIC, such as low power and high performance and still offer more flexible implementations \cite{fanucci2006asip}. While most of the hardware solutions still use fixed-point arithmetic to meet hardware compactness and low complexity architectures \cite{campos4}, these solutions affect the precision and accuracy of the results while adding additional steps into the design flow, for instance, the floating-point to fixed-point conversion and numerical analysis to minimise the quantisation error. 

This paper presents the hardware acceleration of spatial filters in custom floating-point arithmetic. Examples to motivate the acceleration will range from linear spatial filters such as the Sobel operator to non-linear filters with generic kernels, easily implemented in software at the cost of a very slow computation, making real-time video processing unfeasible. Furthermore, using a domain-specific language will allow the rapid prototyping of image processing algorithms in custom floating-point arithmetic, allowing non-experts to quickly develop real-time video processing applications \cite{campos2021fpga}\cite{campos2022fpga}\cite{De-Sousa-Campos2022}.

The rest of this paper is organised as follows. First, section \ref{sec:related_work} briefly discusses previous work relevant to this topic. Section \ref{sec:spatial_filters_fpga} describes the floating-point architecture of the window generator and the architecture of the linear and non-linear spatial filters, with examples ranging from $3\times 3$ and $5\times 5$ convolutions to median filters and generic filters with custom kernels. Next, section \ref{sec:results} discusses the FPGA implementation results and the achieved throughput of the spatial filters implemented in software and hardware. Section \ref{sec:dsl} presents a domain-specific language (DSL) to generate custom floating-point arithmetic using a syntax similar to Matlab. The DSL, alongside the floating-point library, enables non-experts to develop real-time spatial filters in FPGA quickly. Finally, Section \ref{sec:conclusion} concludes the work.

\section{Related Work}\label{sec:related_work}
Hand-coded hardware description language of complex algorithms, such as non-linear spatial filters, is a  tedious and error-prone task. Many works have focused on code autogeneration for different applications in both FPGA and ASIC. For example, in \cite{schmuland2012cad}, autogeneration of VHDL code for FPGA and ASIC implementations of Fast Fourier Transforms (FFT) is proposed. A software tool written in Matlab receives input parameters such as the FFT size and generates the parallel architecture of the FFT structure, where every butterfly is instantiated, and the twiddle factors are optimized using constant multipliers. In \cite{serre2018dsl}, an FFT hardware generator is implemented in Scala. The generator's input is a high-level description of an FFT algorithm. The output is a synchronized design written in Verilog. The generator employs three layers of Domain-Specific Languages (DSL) to represent and optimize the FFT at different levels of abstraction.

In \cite{weinstein2007methodology}, an auto-generation toolkit for the development of neural models in FPGA is proposed. This design tool enables model construction-level alterations (e.g., adjustment of model population size or insertion/deletion of an ionic conductance). The resulting implemented model performs at a theoretical 8.7 real-time utilizing 90\% of logic elements within a Xilinx Virtex-4 XC4VSX35-fg676-10 FPGA. A large portion of the infrastructure can be readily autogenerated, freeing the modeller to focus on the modelling task. A variety of tools were generated in Matlab, focusing on three particular components: the state subsystems, the parameter subsystem, and an output subsystem to interface the analogic outputs. These subsystems are created as dynamically linked Simulink subsystems, such that when the database is altered, a simple command will automatically update the library blocks and the affected models.

The work in \cite{mullapudi2015polymage} presents the design and implementation of PolyMage, a domain-specific language and compiler for image processing pipelines. An image processing pipeline can be
viewed as a graph of interconnected stages which process images successively with operations on image pixels. The traditional options for applications that demand high performance are to use optimized libraries like OpenCV or to optimize manually. While using libraries precludes optimization across library routines, manual optimization accounting for both parallelism and locality is very tedious. PolyMage automatically
generates high-performance implementations of image processing pipelines expressed in a high-level declarative language, taking a high-level specification as input and automatically transforming it into a high-performance parallel implementation.

Halide is a domain-specific language for image processing pipelines that automates the code generation of high-performance image processing algorithms targetting CPUs and GPUs \cite{ragan2013halide}. In \cite{li2020heterohalide}, the Halide DSL is extended to map FPGA implementations of those pipelines using the HeteroHalide.

In \cite{zhao2019automatic}, a framework called Tomato, designed to automate the process of generating efficient CNN accelerators, is presented. The auto-generated design is pipelined, and each convolution layer uses multi-precision and multi-arithmetic. The workflow can quantize weights to both fixed-point and shift values at different precisions and keeps activations to fixed-point numbers. In addition, it transforms batch normalization into simple affine operations with fixed-point scaling and offset factors. The framework utilizes the Roll-Unrolled compute pattern in hardware and provides flexibility in rolling computations in the channel dimension. As a result, the guided rolling reduces computing while keeping the input stream stall-free. The results showed state-of-the-art model accuracy, latency, and throughput performance. The implemented accelerator for MobileNet is fully pipelined with sub-millisecond latency (0.32ms) and can classify at around 3000 frames per second.

In \cite{reiche2014code}, the Heterogeneous Image Processing Acceleration (HIPAcc) framework was proposed as a means for automatic code generation of image processing algorithms written in C-based HLS from a high-level DSL description targeting FPGAs. The HIPAcc compiler is based on the Clang/LLVM 3.4 compiler infrastructure. The proposed compiler parses C/C ++ code and generates an internal Abstract Syntax Tree (AST) representation utilizing the Clang front end. In addition, the DSL contains language components that capture the computational algorithm operations. The most important components are Images, which are objects that represent two-dimensional data types. In addition, local operators employ methods to access (read or write) images. Also, a sliding window to iterate over neighbouring pixels and border handling to deal with edges resulted from convolution operators \cite{membarth2015hipa}\cite{reiche2017generating}.

\section{FPGA Implementation of Spatial Filters}\label{sec:spatial_filters_fpga}
This section describes the proposed hardware implementations for the filter architectures. Subsection \ref{sec:filter_structure} disscuses the general strucute of the spatial filters. Subsection \ref{sec:linear_convolution} discusses linear convolutions. Subsections \ref{sec:median} and \ref{subsec:generic_filter} examine the implementation of median filters and a complex non-linear spatial filter. 

\subsection{Generic Filter Structure}\label{sec:filter_structure}
The generic structure of a 2D convolution with kernel dimensions of height and width $H$ and $W$, respectively, requires $H\times W$ registers to store the pixels inside a sliding window. Due to the nature of the pixel stream, one pixel is inputted into each window row at each clock cycle and shifted horizontally with the flip-flops. A kernel of height $H$ requires $H-1$ line buffers to cache video lines, reorder the raster scan, and form a bi-dimensional window at each clock cycle. This approach leads to massive memory savings by streaming the image into the window rather than processing the whole image frame, allowing the direct stream of cameras or other HDMI sources to the FPGA system in real-time \cite{bailey2011image}. 

\begin{figure}[h]
    \centering 
    \includegraphics[width=\linewidth]{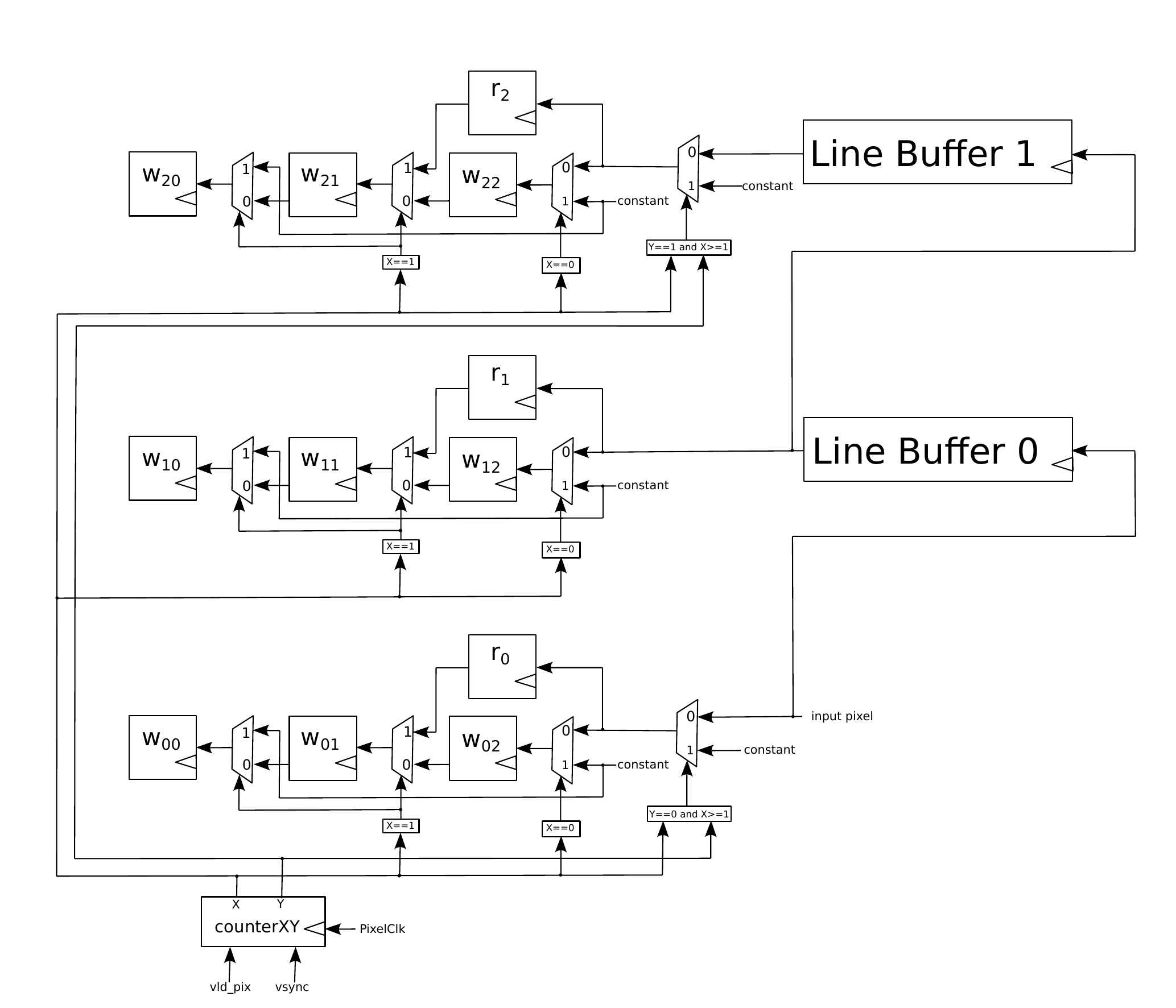}
    \caption{Window generation structure for a filter of dimension $3\times 3$}
    \label{fig:window_3x3}
\end{figure}

Figure \ref{fig:window_3x3} shows the filter structure of a spatial filter of dimension $3\times 3$; as shown in the diagram, two-line buffers inferring dual-port RAM cache the video lines. Nine registers process the pixel neighbourhood that is convolved with the pixel stream. In addition to that, three registers $r_0$, $r_1$ and $r_2$, are used to manipulate border handling. If those registers are ignored, the convolution produces incorrect results at the beginning and end of each line and frame. Introducing those registers makes temporal copies of the pixels inside the window to either replicate delayed pixel values or pass a constant value. Additional muxes are required to control the border handling, selecting a delayed pixel or a pre-defined value. Different techniques can be used, for instance, constant extension, mirroring or reflection \cite{bailey2011image}. An additional $\frac{H\times (W-1)}{2}$ registers and $H\times (W+1)-1$ muxes are required to control the border management for a window of dimensions $H\times W$. Temporal controllers are also required to trigger the muxes when the current pixel is located at the frame edges. Those controllers use sequential counters synchronised with the video signals to control the write address of the dual-port RAMs that cache the video lines. 

\begin{figure}[h]
    \centering 
    \includegraphics[width=\linewidth]{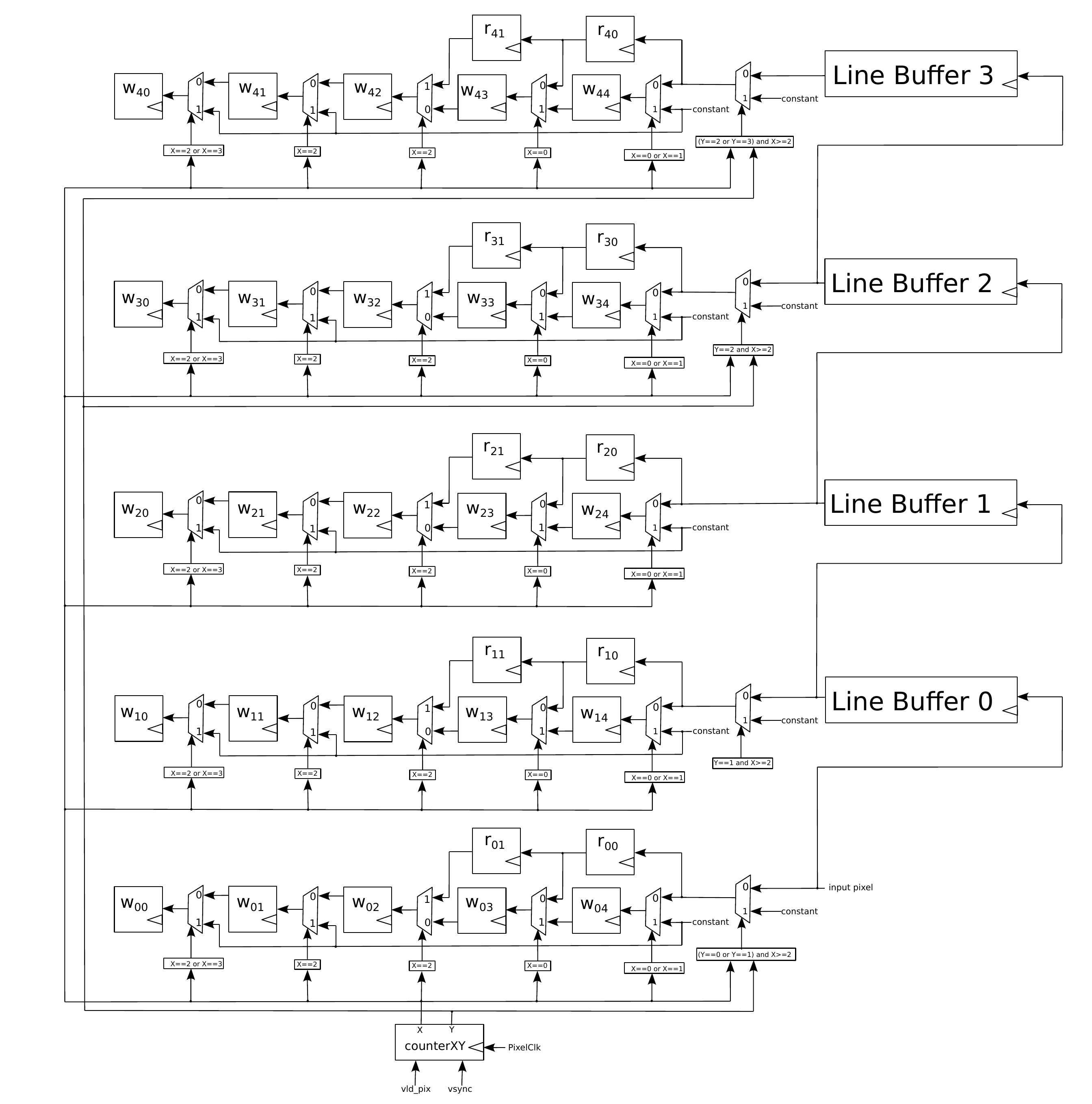}
    \caption{Window generation structure for a filter of dimension $5\times 5$}
    \label{fig:window_5x5}
\end{figure}

Figure \ref{fig:window_5x5} shows the structure for generating a window neighbourhood of dimensions $5\times 5$. For this case, in addition to the 25 registers to store the $5\times 5$ window neighbourhood, 10 temporal registers control the border handling and four line buffers are required to store video lines.

\begin{figure}[h]
    \centering 
    \includegraphics[width=0.75\linewidth]{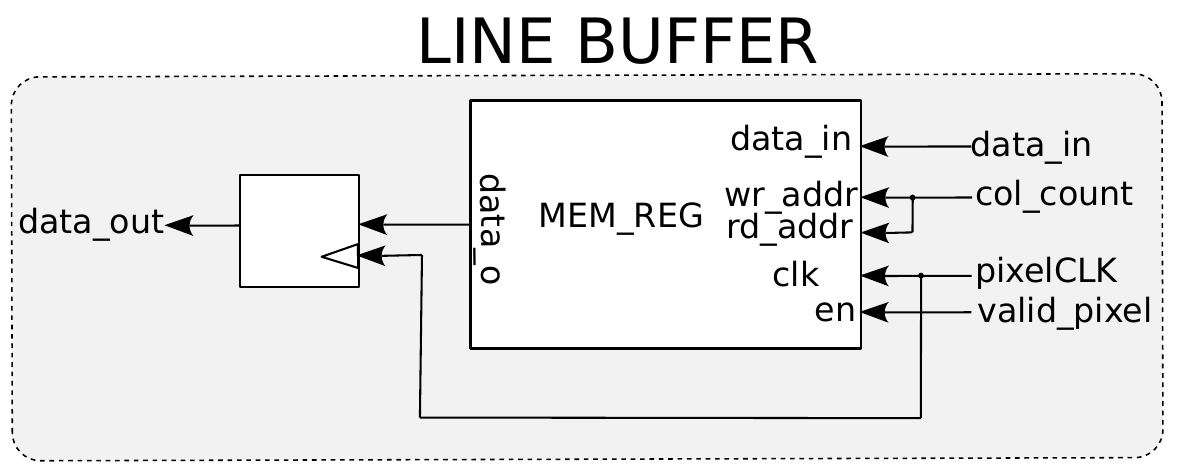}
    \caption{Line buffer memory makes inference of dual-port RAM}
    \label{fig:line_buffer}
\end{figure}

The implementation of the line buffers efficiently makes inference of dual-port RAMs. This memory structure, illustrated in figure \ref{fig:line_buffer}, is composed of blocks of 36 Kb with two independent access ports for reading and writing data, allowing read and write data at the same address in the same clock cycle \cite{mehta2011xilinx}. 

The window generator writes the current line reading data from the previous line, acting as a circular FIFO updated at every line, with the write enable of the dual-port RAM connected to the valid pixel signal of the video interface, bypassing blanking pixels. Furthermore, the inference of the block RAM requires one additional flip-flop to write the data, leading to time misalignment in the window generator structure. One solution to avoid this temporal misalignment is to design the dual-port RAM to read data at the clock's positive edge and write at the negative edge. 

\subsection{Linear Convolution}\label{sec:linear_convolution}
The linear convolution is the most common set of spatial filters in image processing and CNN applications. The filter function operates in a window neighbourhood during each clock cycle. Every pixel inside the window is multiplied by a kernel coefficient, and all resulted products are summed to compute the output pixel. If the kernel coefficients are constant and the data is represented in fixed-point format, then the operation is usually implemented using a multiplier-less approach. However, suppose the kernel values are not constant, and the data representation needs high precision fixed-point or floating-point format. In that case, DSP blocks are usually required to perform the multiplications in the window filter. 

\begin{figure}[h]
    \centering 
    \includegraphics[width=\linewidth]{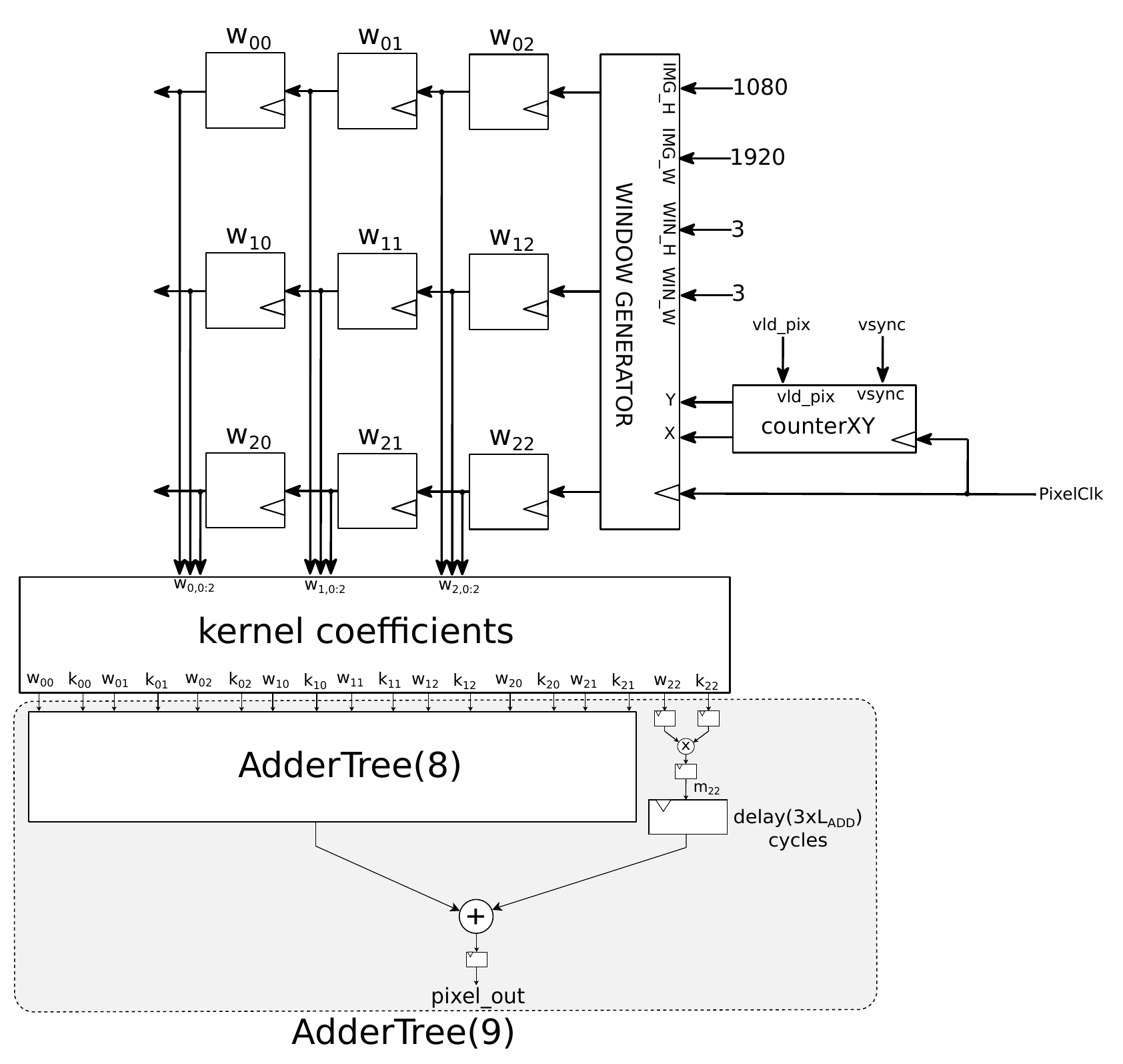}
    \caption{2D convolution with kernel dimensions 3x3}
    \label{fig:conv3x3}
\end{figure}

Another essential factor to consider is the structure of the adder trees, which are responsible for accumulating the summation in the window neighbourhood. Each resulted product is accumulated to provide the final result. For example, figure \ref{fig:conv3x3} illustrates the convolution of dimensions $3\times 3$. Nine pixels are multiplied by each kernel coefficient, resulting in nine product operations. The adder tree should then accumulate each coefficient in pairs to compute the final result. High-performance implementations must be pipelined in stages to synchronise data and avoid pixel flickering, which is the oscillation in the light intensity of a pixel due to unstable values. 

\begin{figure}[h]
    \centering 
    \includegraphics[width=0.75\linewidth]{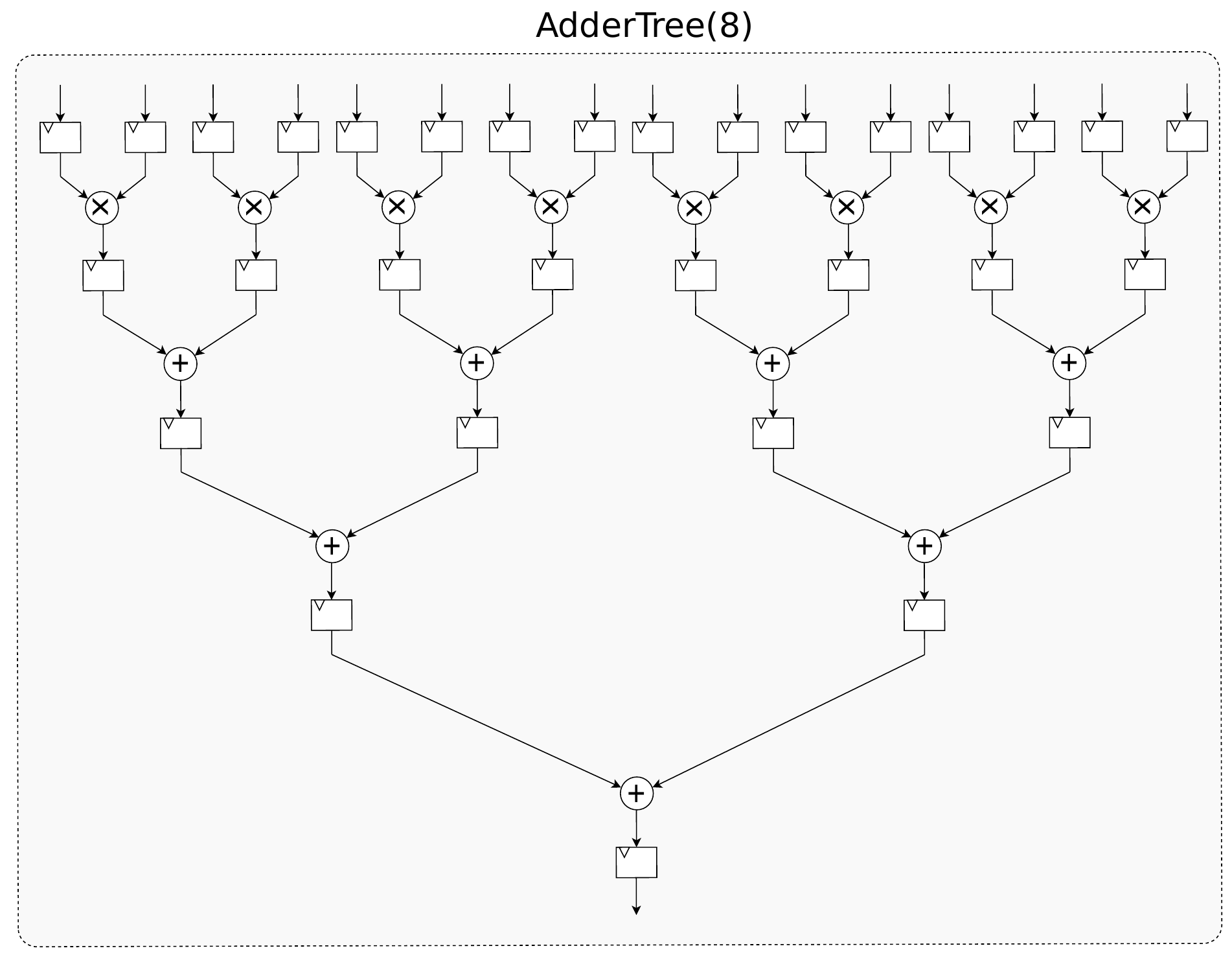}
    \caption{A 3-stage pipeline adder tree of eight inputs}
    \label{fig:adder_tree}
\end{figure}

The computation of the output pixel in figure \ref{fig:conv3x3} is done with the sum of the result of the adder tree illustrated in figure \ref{fig:adder_tree} ($AdderTree(8)$) with the result of the multiplication of the ninth-pixel pixel $w_{22}$ by its respective coefficient $k_{22}$. 

$AdderTree(8)$ is a 3-stage pipeline of eight adders\footnote{The number of stages of the adder tree is $N_s = \lfloor log_2(N_i) \rfloor$, where $N_i$ is the number of inputs in the adder tree.}. In the first stage, four adders are accumulated in parallel. Each subsequent stage sums up the results of the previous stages. The output of the third stage completes after a sequence of three parallel additions. If the addition is implemented in fixed-point format, it usually takes one clock cycle to compute the operation. Floating-point implementations have additional latencies\footnote{In the experiments conducted in this work, the floating-point adder consumes six clock cycles to compute the first operation and has a throughput of one operation per cycle.}. Therefore, to compute the output pixel, $m_{22}$ needs to be delayed by $3\times L_{ADD}$ clock cycles, where $L_{ADD}$ is the latency of the floating-point addition. The total latency of the $AdderTree(9)$ is $4\times L_{ADD}$.

\begin{figure}[h]
    \centering 
    \includegraphics[width=\linewidth]{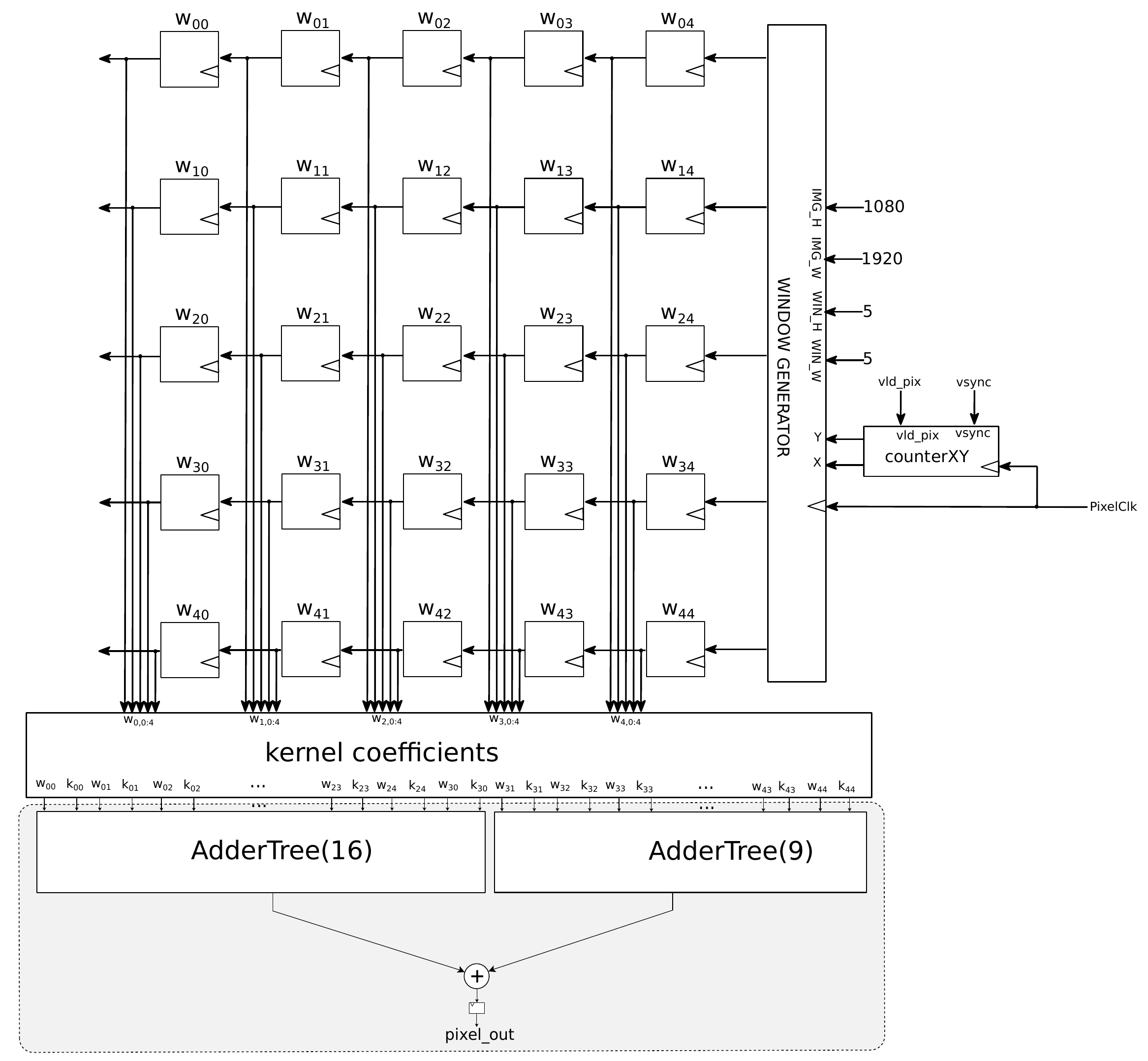}
    \caption{2D convolution with kernel dimensions $5\times 5$}
    \label{fig:conv5x5}
\end{figure}

The adder tree structure for a $5 \times 5$ linear convolution is shown in figure \ref{fig:conv5x5}. Similarly, 25 pixels must be accumulated after multiplication with the kernel coefficients. 

As a design decision rule, let $AdderTree(N_{add})$ be an adder tree of $N_{add}=N_0+N_1$ inputs, which has a latency of $L_{ADD}\times \left \lceil{log_2(N_{add})}\right \rceil$. $AdderTree(N_{add})$ can be formed by the addition of two adder trees of order $N_0$ and $N_1$. $N_0=2^{\left \lfloor{log_2(N_{add})}\right \rfloor}$ is the closest power of two smaller than $N_{add}$. If $N_1$ is not a power of two, $AdderTree(N_1)$ will be decomposed recursively into two adder trees.

In the particular case of 25 additions, the adder tree can be composed of 16 additions (the $AdderTree(16)$ in figure \ref{fig:conv5x5}) that will take $4\times L_{ADD}$ cycles to execute the first operation. Then, the remaining nine additions are computed using an ($AdderTree(9)$),  which is comprised of $AdderTree(8)$ and $AdderTree(1)$, as stated previously.

\subsection{Median Filter}\label{sec:median}
\begin{figure}[h]
    \centering 
    \includegraphics[width=\linewidth]{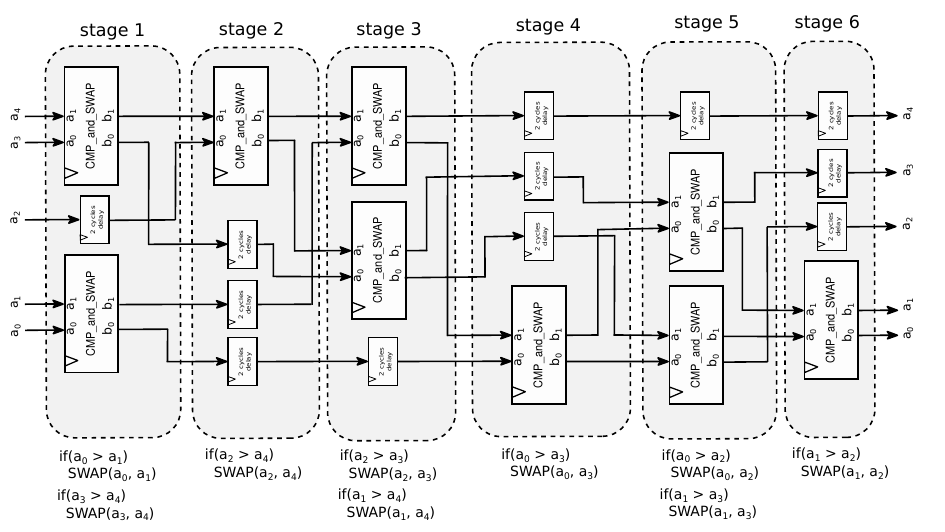}
    \caption{Bose-Nelson Sorting network for sorting five inputs}
    \label{fig:sort_network}
\end{figure}

The median filter is one of the most common non-linear spatial filters for image denoising.  Unlike the convolution, which operates additions and multiplications inside the window neighbourhood, the median filter sorts the pixels surrounding a window region.  The middle pixel of the sorted elements is the output of the median filter.  Typical hardware implementations of median filters consist of the window generation similar to the linear convolution that uses line buffers to store lines from the streaming video.  The arithmetic computation of the filter uses a sorting network to sort a footprint array of pixels.  The footprint is the whole matrix containing all the elements in the window or a part of the region selecting specific pixels.  Popular techniques for median filters are sorting networks based on Batcher's or Bose-Nelson's algorithms.  These algorithms are based on a series of swap and compare operations.  Figure \ref{fig:sort_network} shows the steps for sorting five pixels inside a window of $3\times 3$ dimensions. The Bose-Nelson algorithm can compute the sorting operation using nine $CMP\_and\_SWAP$ computations.  The sorting network is parallelised in six pipelined stages.  Each $CMP\_and\_SWAP$ operation compares a pair of elements $a_i$ and $a_j$, and if $a_i > a_j$, the pair $[a_i, a_j]$ is swapped ($[a_i, a_j] \leftarrow [a_j, a_i]$).

\begin{figure}[h]
    \centering 
    \includegraphics[width=\linewidth]{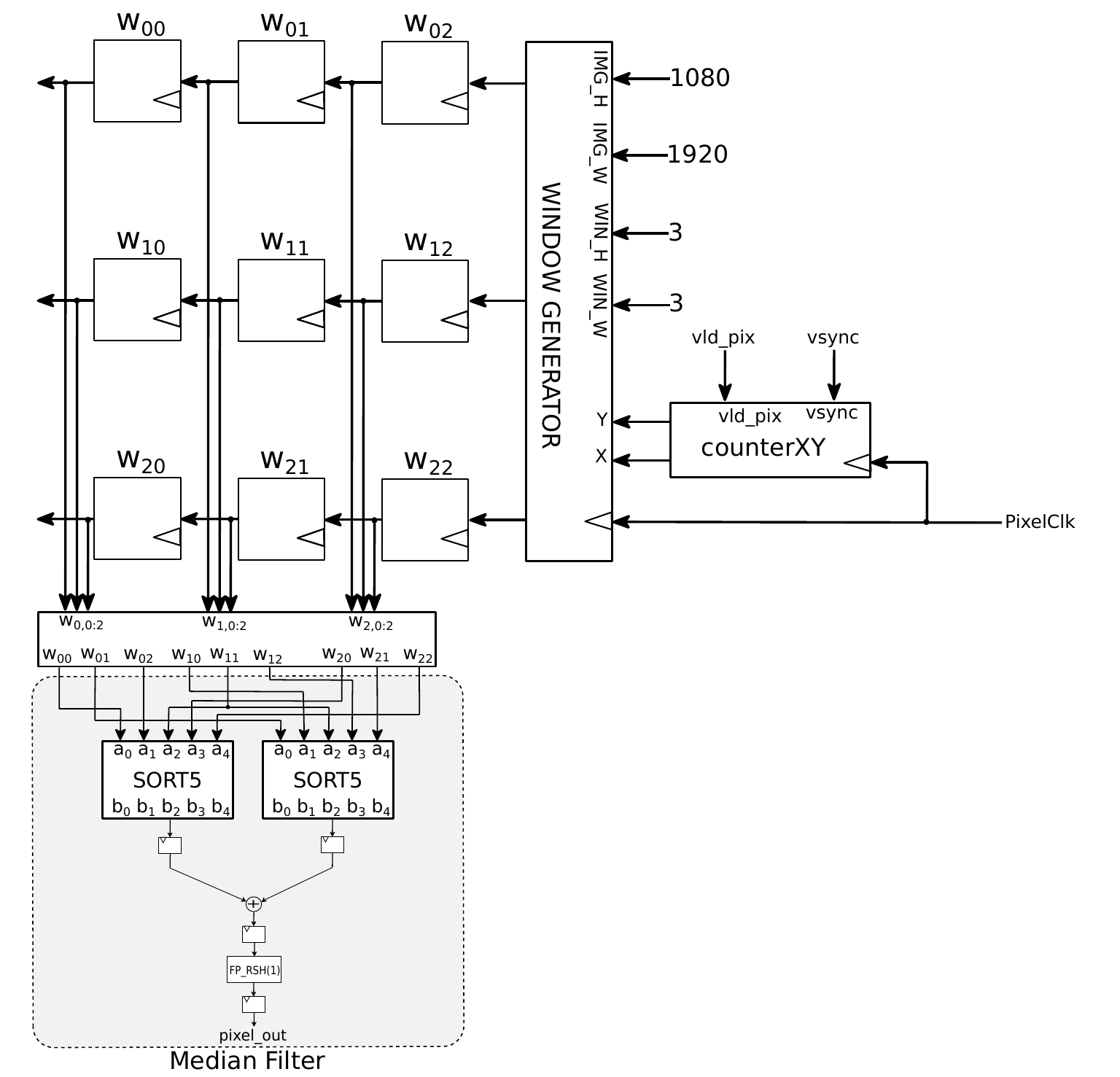}
    \caption{The hardware architecture of a floating-point median filter contains two sorting networks that operate according to the circuit illustrated in \ref{fig:sort_network}. The output pixel is the mean of the two $SORT_5$ that operates in parallel.}
    \label{fig:median_filter}
\end{figure}

Figure \ref{fig:median_filter} shows the hardware architecture of a median filter that is composed of two $SORT_5$ networks. The output pixel is the mean of the two $SORT_5$ blocks that operate a set of five pixels, each $SORT_5$ operating in parallel. The window generator creates a vector $w$ of floating-point pixels in each clock cycle. The $SORT_5$ circuit on the right in figure \ref{fig:median_filter} selects the pixels $a_0=w_{01}$, $a_1=w_{10}$, $a_2=w_{11}$, $a_3=w_{12}$ and $a_4=w_{21}$. At the first stage of the $SORT_5$, two $CMP\_and\_SWAP$ operate on the pairs $a_0$ and $a_1$, $a_3$ and $a_4$. Because the $CMP\_and\_SWAP$ takes two clock cycles to operate, $a_4$ needs to be delayed by two clock cycles. During the second stage, a $CMP\_and\_SWAP$ operates in the pixels $a_2$ and $a_4$, while the pixels $a_0$, $a_1$ and $a_3$ are delayed by two clock cycles. This process repeats in the subsequent stages to compute a $SORT_5$, taking a total latency of 12 cycles. The output of the sorting network is $a_2$. The median filter computes the sum of both sorting networks that run concurrently\footnote{The $SORT_5$ circuit on the left select the pixels  $a_0=w_{00}$, $a_1=w_{02}$, $a_2=w_{11}$, $a_3=w_{20}$ and $a_4=w_{22}$.}. Finally, this result is divided by two with a floating-point right-shift operation\footnote{If $x$ is a fixed-point number, right-shifting $N$ bits of $x$ results in $y=\frac{x}{2^N}$. If $x$ is a floating-point number, $y$ is computed with the subtraction of $N$ from the exponent of $x$. Similarly, left-shifting $N$ bits of $x$ results in $z=x \times 2^N$ for $x$ and $z$ in fixed-point. In floating-point, $z$ is computed with the addition of $N$ to the exponent of $x$.}. 

\subsection{Spatial Filters with Generic Functions}\label{subsec:generic_filter}

So far, the spatial filters covered are composed of a pre-defined function that operates in a window region. For example, the linear convolution can be defined by a function $conv_{H\times W}(w, k)$, where $w$ and $k$ are matrixes of dimensions $H$ and $W$, as can be seen in equation \ref{eq:lin_conv}. This function is computed using the adder tree structures already described in subsection \ref{sec:linear_convolution}. On the other hand, the median filter from subsection \ref{sec:median} operates in a window $w_i$ and produces a sorted window $w_o = SORT_N(w_i)$. The output of the median is the central pixel of the sorted matrix $w_o$ (the element at the position $(\frac{H-1}{2}, \frac{W-1}{2})$)\footnote{The median filter from subsection \ref{sec:median} is composed of two $SORT_5$ operations instead of one $SORT_9$. This was a design decision justified with the fact that two $SORT_5$ operation can be defined with less swap and compare operations that one $SORT_9$.}. The Python library \textit{scipy} provides the optimised functions \textit{convolve2d} and \textit{median} to compute 2D convolution and median filters, respectively \cite{bressert2012scipy}.

\begin{equation}\label{eq:lin_conv}
\resizebox{\hsize}{!}{$
conv_{H\times W}(w,k) =  
w_{H\times W} * k_{H\times W} = 
\begin{bmatrix}
w_{00} &\hdots  &w_{0W}  \\
\vdots &\ddots  &  \vdots\\
w_{H0}&  \hdots &w_{HW}  \\
\end{bmatrix} 
*
\begin{bmatrix}
k_{00} &\hdots  &k_{0W}  \\
\vdots &\ddots  &  \vdots\\
k_{H0}&  \hdots &k_{HW}  \\\end{bmatrix} = \sum \limits_{i=0}^{H-1}\sum \limits_{j=0}^{W-1}w_{ij}\times k_{ij}  
$}
\end{equation}

Suppose the spatial filter needs to be defined with a generic function; for instance, the function $f^{\zeta}$ in equation \ref{eq:nl_funcs}. The software implementation of those non-linear filters is straightforward. Matlab provides the function \textit{nlfilter}, which operates in a window $w$ using a custom defined function \cite{matlab_nlfilter}. However, the ease of implementation of those functions in software comes at the cost of a very slow computation, making the video processing in real-time unfeasible. 

\begin{equation}\label{eq:nl_funcs}
\resizebox{\hsize}{!}{$
\left\{\begin{matrix}
\begin{bmatrix} 
f^{\alpha}\\ 
f^{\beta}\\ 
f^{\delta}\end{bmatrix} 
= 
\begin{bmatrix} 
0.5\times(\sqrt{max(w_{00},1)\times max(w_{02},1)} + \sqrt{max(w_{20},1)\times max(w_{22},1)})\\ 
8\times(log_2(max(w_{01},1)\times max(w_{21},1)) + log_2(max(w_{10},1)\times max(w_{12},1)))\\
2^{0.0313\times max(w_{11},1)}
\end{bmatrix} \\

f^{\zeta} = 
\left\{\begin{matrix}
f^{\alpha}\times \left(\frac{f^{\delta}}{f^{\beta}}\right) & \mbox{, if $f^{\beta} > f^{\delta}$}\\ 
\\
f^{\alpha}\times \left(\frac{f^{\beta}}{f^{\delta}}\right) & \mbox{, otherwise} 
\end{matrix}\right.
\end{matrix}\right.
$}
\end{equation}

\begin{figure}[h]
    \centering 
    \includegraphics[width=\linewidth]{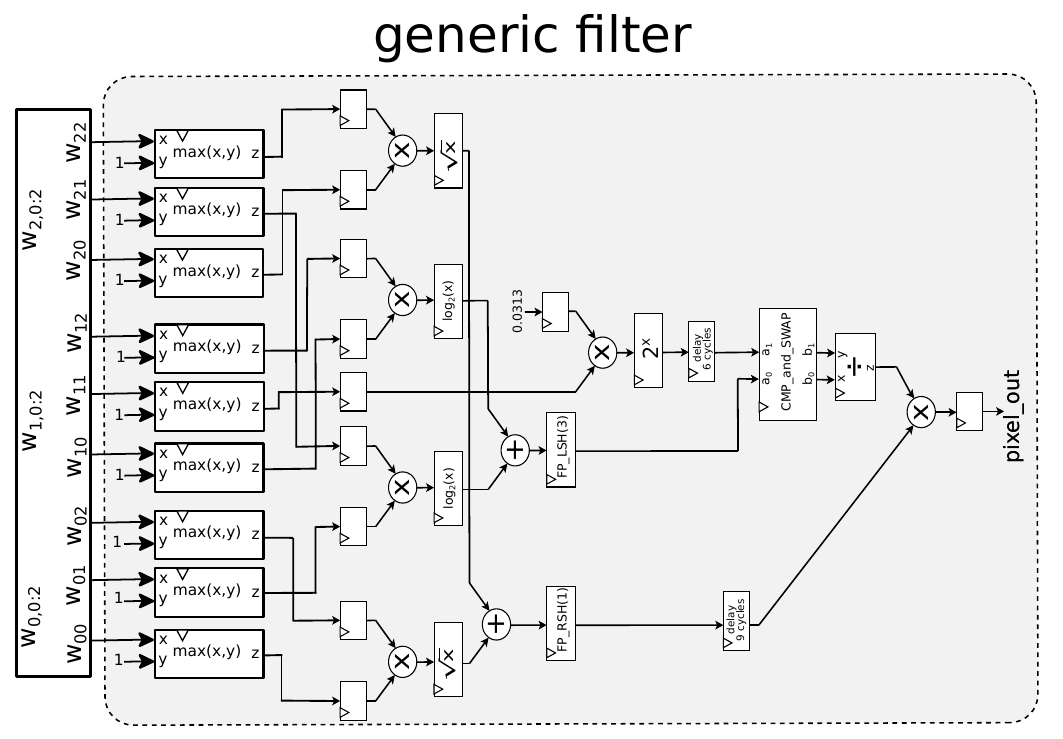}
    \caption{Hardware architecture of a non-linear floating-point spatial filter with a custom function defined in equation \ref{eq:nl_funcs}.}
    \label{fig:nl_filter}
\end{figure}

The spatial filter structure with the custom function defined in equation \ref{eq:nl_funcs} is shown in figure \ref{fig:nl_filter}. The output of the generic filter is the function $f^{\zeta}$, that is a composite function and has three input functions $f^{\alpha}$, $f^{\beta}$ and $f^{\delta}$. Note that $f^{\zeta}$ can be written in the form $f^{\zeta} = f^{\alpha} \times f^{\phi}$, with $f^{\phi} = \frac{f^{\beta'}}{f^{\delta'}}$ and  $[f^{\beta'},f^{\delta'}] = CMP\_and\_SWAP(f^{\beta},f^{\delta})$. Here, the function $CMP\_and\_SWAP$ is the same function used to build the $SORT_5$ network used in the median filter.

\begin{figure}[h]
    \centering 
    \includegraphics[width=\linewidth]{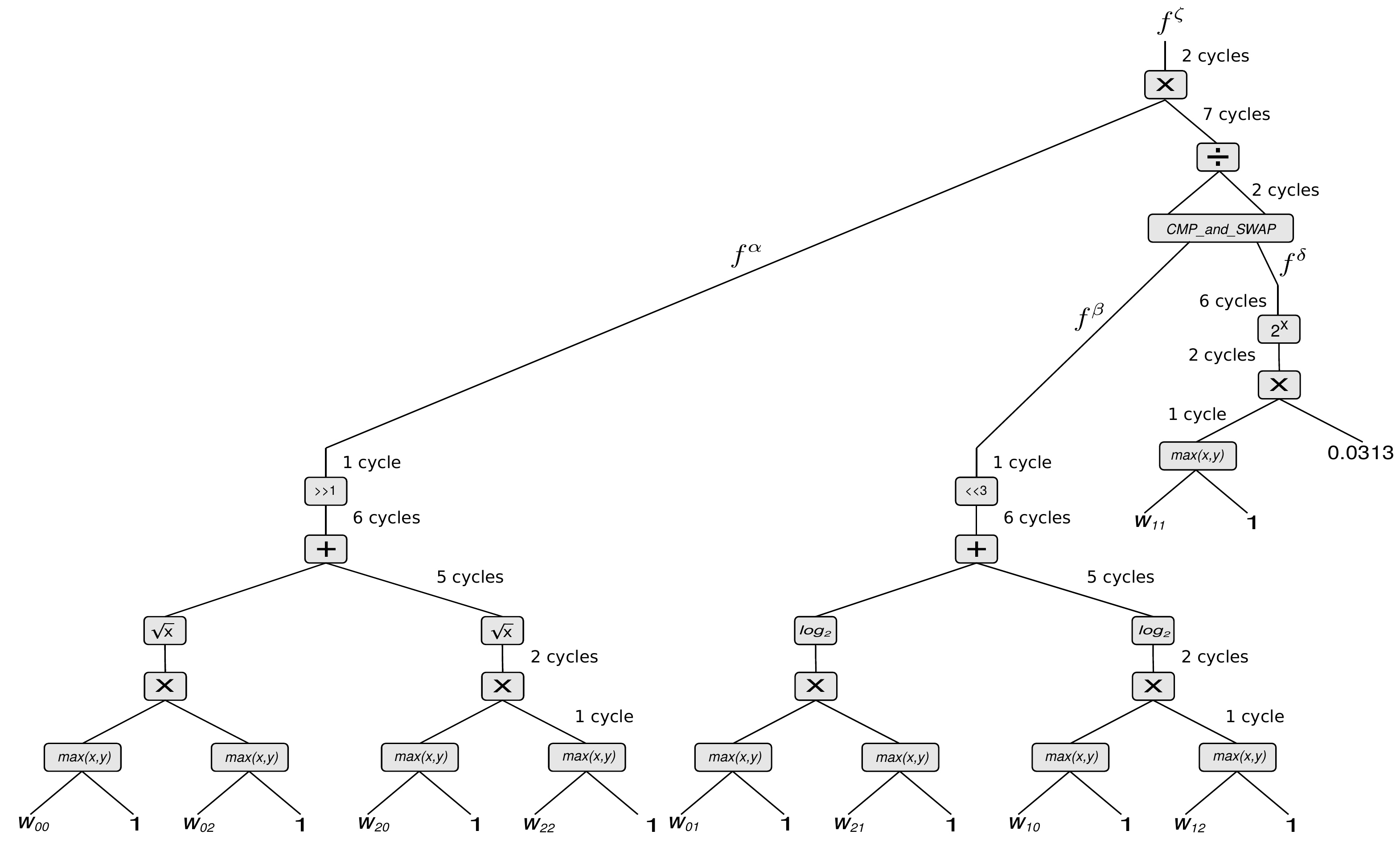}
    \caption{Abstract Syntax Tree (AST) for the steps to compute the function $f^{\zeta}$ in equation \ref{eq:nl_funcs}}
    \label{fig:ast}
\end{figure}

Figure \ref{fig:ast} shows the Abstract Syntax Tree (AST) that represents the steps to compute the function $f^{\zeta}$ in equation \ref{eq:nl_funcs}. It is essential to observe in figure \ref{fig:nl_filter} that $f^{\delta}$ is delayed by six clock cycles before it is inputted into the function $CMP\_and\_SWAP$. This happens because the latency of $f^{\beta}$ is 15 cycles, whereas the latency of $f^{\delta}$ is nine cycles. 

The latency of $f^{\alpha}$ can be computed as follows from the analysis of the AST in figure \ref{fig:ast}:\footnote{Similar steps can be used to compute the latency of $f^{\delta}$.}

\begin{enumerate}
\item compute the maximum of the inputs $w_{00}$, $w_{02}$, $w_{20}$ and $w_{22}$  with 1;\footnote{The operation $w'_{ij}=max(w_{ij,},1)$ has a latency of one clock cycle.}
\item compute the operations $w'_{00}\times w'_{02}$ and $w'_{20}\times w'_{22}$;\footnote{The operation of multiplication has a latency of two clock cycles.}
\item for each result of the multiplication, compute its square root\footnote{Here, the square root is computed using a four segments, degree-2 polynomial approximation and has a latency of five clock cycles.};
\item add the result of the two square roots\footnote{The adder's latency is six clock cycles.};
\item the multiplication by $0.5$ (a constant that is a power of two) can be computed using a floating-point right-shifter with one cycle latency;
\item the latency of the function $f^{\alpha}$ is the sum of the latencies in each step to compute $f^{\alpha}$.\footnote{Note from figure \ref{fig:ast} that the AST of $f^{\alpha}$ and $f^{\beta}$ only differ in the step of the addition, that sums the outputs of two $log_2$ operations instead of two square-roots. However, both operations have a latency of five cycles.}
\end{enumerate}

After $f^{\beta}$ and $f^{\delta}$ are aligned to have the same latencies\footnote{At this stage, $f^{\delta}$ was already delayed by six clock cycles to match the latency of $f^{\beta}$. Another two clock cycles are required to compute the operation $CMP\_and\_SWAP$ and produce the ouptus $f^{\beta'}$ and $f^{\delta'}$.}, a division that consumes seven clock cycles\footnote{Here, the division is computed using a four segments, degree-3 polynomial approximation.} operate in $f^{\beta'}$ and $f^{\delta'}$ to produce the result $f^{\phi}$. At this point, the latency of $f^{\phi}$ is 24 cycles, which requires $f^{\alpha}$ to be delayed by nine cycles before being multiplied by $f^{\alpha}$ to compute the final result $f^{\zeta}$.

A general formula to compute the delays of signals to match latencies can be defined as follows. Let $\lambda(s_i)$ and $\lambda(s_j)$ be the latencies of signals $s_i$ and $s_j$ after operations $\Theta_i$ and $\Theta_j$, respectively. If $s_i$ and $s_j$ are the inputs of another operator $\Theta_{ij}$ and $\lambda(s_i) \neq \lambda(s_j)$, then $\lambda(s_i)$ and $\lambda(s_j)$ need to be updated such that $\lambda(s_i) = \lambda(s_j)$. 

Let also $\lambda(s_{i+1})$ and $\lambda(s_{j+1})$ be the updated latencies of signals $s_i$ and $s_j$ when $s_i$ and $s_j$ are ouputs of $\Theta_i$ and $\Theta_j$, respectively, and also inputs of $\Theta_{ij}$. The updated latencies can be computed according to the expression $\lambda(s_{i+1}) = max(\lambda(s_i), \lambda(s_i)) = \lambda(s_{j+1})$. Yet, the number of cycles to delay signal $s_i$ such that $\lambda(s_i) =\lambda(s_j)$ is $\Delta(s_i,s_j) = \lambda(s_{i+1})-\lambda(s_i)$. 

\section{Results and discussions}\label{sec:results}
This section presents an analysis of the spatial filters introduced in section \ref{sec:spatial_filters_fpga}. Subsection \ref{subsec:filter_fps} discusses the achieved frame rate of the spatial filters from figures \ref{fig:conv3x3}, \ref{fig:conv5x5}, \ref{fig:median_filter} and \ref{fig:nl_filter} for both implementations in software and hardware. Subsection \ref{subsec:filter_resources} shows the FPGA implementation results of the spatial filters using a Zybo Z7-20 FPGA.

\subsection{Achieved Throughput for the Spatial Filters implemented in software and hardware}\label{subsec:filter_fps}
There is a tradeoff between ease of implementation and computational performance when comparing hardware and software algorithms in image processing. Hardware implementations often lead to a high code density and complex algorithms. On the other hand, software libraries are usually available, providing functions for convolutions and spatial filters with few lines of code. However, as already stated, this development facility compromises the real-time processing of vision applications.

\begin{table}[h]
\centering
\begin{tabular}{|c|c|c|c|c|} 
\hline
\multicolumn{2}{|c|}{\multirow{2}{*}{\textbf{Filter Type}}} & \multicolumn{3}{c|}{\textbf{Frame Resolution}}                                                               \\ 
\cline{3-5}
\multicolumn{2}{|c|}{}                                      & $640\times 480$                              & $1280\times 720$                          & $1920\times 1080$                         \\ 
\hline
\hline
\multirow{4}{*}{\rot{Software}} & \textit{conv}$_{3x3}$       & 295.71 FPS                  & 67.34 FPS                & 34.22 FPS                \\ 
\cline{2-5}
                                   & \textit{conv}$_{5x5}$       & 162.50 FPS                  & 56.05 FPS                & 22.94 FPS                \\ 
\cline{2-5}
                                   & \textit{median}        & 57.23 FPS                   & 16.58 FPS                & 6.24 FPS                 \\ 
\cline{2-5}
                                   & \textit{nlfilter}      & 0.462 FPS                   & 0.157 FPS                & 0.074 FPS                \\ 
\hline
\hline
\multirow{4}{*}{\rot{Hardware}} & \textit{conv}$_{3x3}$       & \multirow{4}{*}{353.57 FPS} & \multirow{4}{*}{120 FPS} & \multirow{4}{*}{60 FPS}  \\ 
\cline{2-2}
                                   & \textit{conv}$_{5x5}$       &                                      &                                   &                                   \\ 
\cline{2-2}
                                   & \textit{median}        &                                      &                                   &                                   \\ 
\cline{2-2}
                                   & \textit{nlfilter}      &                                      &                                   &                                   \\
\hline
\end{tabular}
\caption{Frame rate of filter functions vs image resolution}\label{tab:filter_fps}
\end{table}

The spatial filters of figures \ref{fig:conv3x3}, \ref{fig:conv5x5}, \ref{fig:median_filter} and \ref{fig:nl_filter} were implemented using the \textit{scipy} library from Python on a Core-i7 computer configuration running at 2.6 GHz. Table \ref{tab:filter_fps} shows the achievable frame rate of the filters  \textit{conv}$_{3x3}$,  \textit{conv}$_{5x5}$, \textit{median} filter  and \textit{nlfilter} (the non-linear filter with the custom function of equation \ref{eq:nl_funcs}). As can be seen from the table, the frame rate is proportional to the frame resolution as well as the kernel size. For instance, the linear convolution can process up to approximately 295.71 FPS at the resolution $640\times 480$ for a kernel size of $3\times 3$. Higher resolutions achieve almost 23 FPS for a kernel size of $5\times 5$ and resolution of 1080p. On the other hand, the median filter is slow compared to linear convolution, achieving roughly six FPS at 1080p, mostly due to the complexity of the algorithms for sorting the sliding windows that streams throughout the image. The frame rate drops drastically for the \textit{nlfilter}, which computes frames of 480p at 0.462 FPS and takes more than 13 seconds to compute a single frame at 1080p resolution. 

\begin{figure*}[h]
    \centering 
    \includegraphics[width=\linewidth]{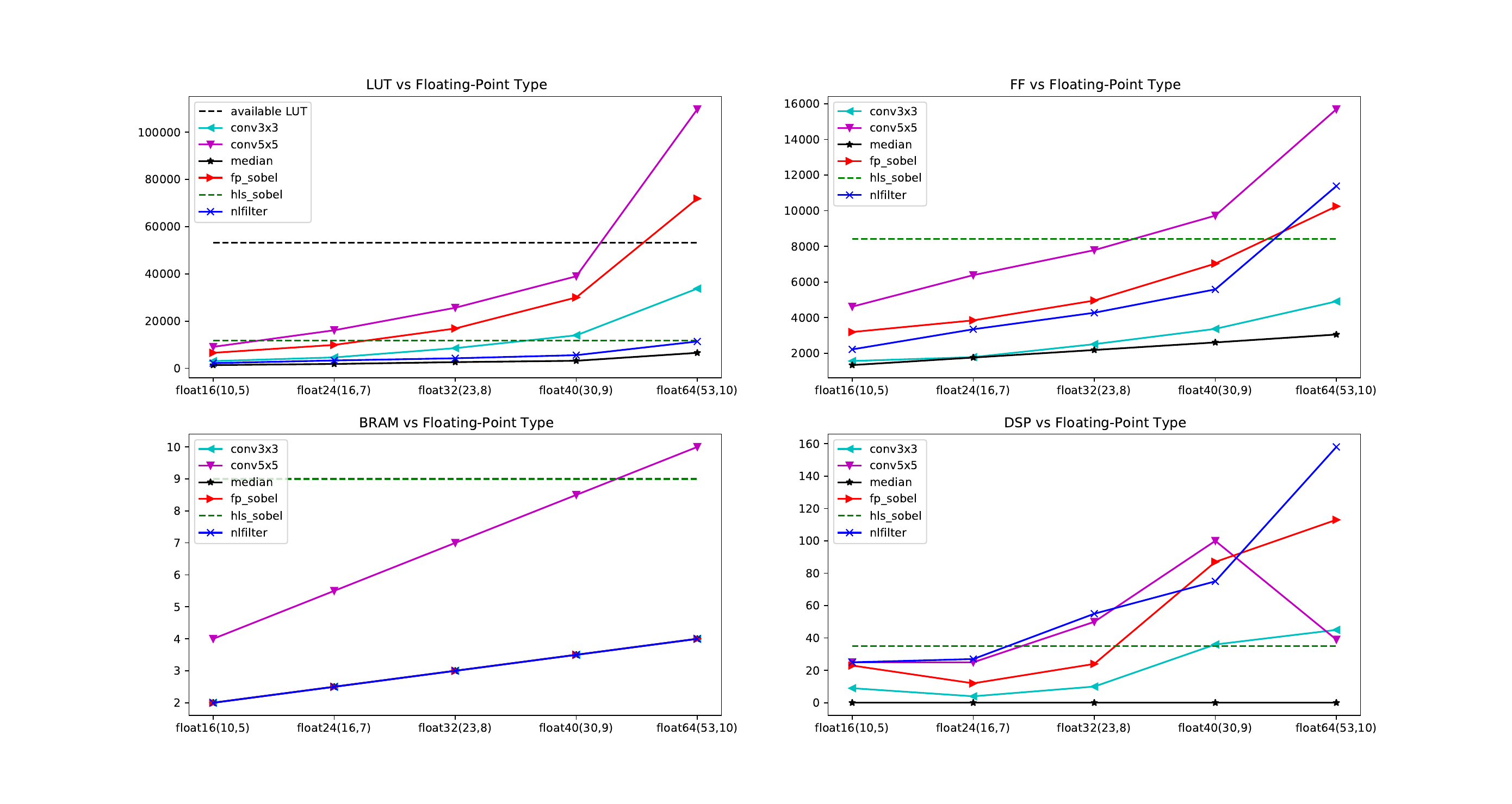}
    \caption{FPGA implementation results for filters vs floating-point type}
    \label{fig:filters_synthesis}
\end{figure*}

The real-time video processing of the filters in table \ref{tab:filter_fps} can be met with implementation in hardware. The Zybo Z7-20 FPGA \cite{zybo-z7-20} board was used for this experiment. Each function is pipelined so that every output pixel is produced every clock cycle. For instance, for $1920\times 1080$ pixels, additional blanking pixels\footnote{There is a total of $2200\times 1125$ pixels resulting from additional 220 blanking lines and 45 blanking pixels in each line. For a frame rate of 60 Hz, a frequency of 148.5 MHz is required for the pixel clock.} account in the frame storage, which provides nearly 6.734 nanoseconds to compute each output pixel from the spatial filters, regardless of the filter function. The pixel clock of 148.5 MHz can operate at 60 FPS at 1080p and 120 FPS and $\approx 353.57$ FPS for 720p and 480p, respectively\footnote{The number of frames per second that a pixel clock can operate in a resolution is given by $FPS = 60\times \frac{148.5}{f_i}$, where $f_i=74.25$ MHz and $f_i=25.2$ MHz for 720p and 480p, respectively.}. 

For images of high resolution, the software implementations of linear convolution achieved up to 57\% of the desired frame rate, whereas the \textit{median} filter accomplished about 10\% of 60 FPS. The hardware acceleration is evident in the function \textit{nlfilter}, which performed about 810 times faster than the software implementation for Full HD images.

\subsection{FPGA Implementation Results for the Spatial Filters}\label{subsec:filter_resources}

Figure \ref{fig:filters_synthesis} shows the FPGA implementation results for six spatial filters. The four spatial filters discussed in subsection \ref{subsec:filter_fps} were implemented in SystemVerilog using custom floating-point arithmetic and synthesized in the Zybo Z7-20 FPGA board. The filters process video of 1080p resolution at 60 Hz. \textit{conv}$_{3x3}$ and \textit{conv}$_{5x5}$ have kernels with reconfigurable coefficients, resulting in DSP block inferences to perform the multiplications required in the inputs of the adder trees. Another spatial filter labelled as \textit{fp\_sobel}  is a SystemVerilog implementation of a Sobel filter, which uses two \textit{conv}$_{3x3}$ filters with kernels $K_x$ and $K_y$ to operate in an image $\Phi_i$ and produce the output $\Phi_o$ according to the expression in equation \ref{eq:fp_sobel}. The five floating-point filters are synthesized with five custom floating-point arithmetics with widths varying from 16 to 64 bits\footnote{The expression $float16(10,5)$ refers to a 16-bit floating-point with 10 bits for the mantissa and 5 bits for the exponent}.

\begin{equation}\label{eq:fp_sobel}
\left\{\begin{matrix}
 K_x = 
\begin{bmatrix}
 1.0&0.0  &-1.0  \\
 2.0&0.0  &-2.0  \\
 1.0&0.0  &-1.0  \\
\end{bmatrix} 

\mbox{; } 
K_y = 
\begin{bmatrix} 
1.0&2.0  &1.0  \\ 
0.0&0.0  &0.0  \\ 
-1.0&-2.0  &-1.0  
\\\end{bmatrix}  \\
 \\
\Phi_o = \sqrt{conv_{3x3}(\Phi_i,K_x)^2+conv_{3x3}(\Phi_i,K_y)^2}
\end{matrix}\right.
\end{equation}

A filter tagged as $hls\_sobel$ was implemented in C++ using high-level synthesis\footnote{The design used the software tool Vivado HLS 2019.1 from Xilinx.} using libraries for image processing from Xilinx to implement the line buffers and compute the convolutions. The implementation used a 24-bit fixed-point to represent the pixel in RGB format. The HLS filter was used as a reference to compare the resource usage of fixed-point implementations with different custom floating-point.

The plots in figure \ref{fig:filters_synthesis} show the resource utilization for LUTs, flip-flops, block RAMs and DSP blocks against the floating-point type\footnote{Note that the HLS filter is not dependent on the floating-point type as it is implemented in 24-bit fixed-point}. LUTs mostly correlate to combinational logic and will be dependent on the complexity of the datapaths. Flip-flops are used to pipeline the datapath to achieve high performance. The datapath complexity depends on the filter complexity, such as the kernel size or the width of the floating-point data. The block RAMs store video lines and are proportional to the frame resolution, the kernel size and the floating-point width. Each floating-point filter uses the window generator structures from figures \ref{fig:window_3x3} and \ref{fig:window_5x5}. The floating-point filters of kernel width $3\times 3$ inferred 2.0 to 4.0 BRAMs when using floating-point widths with 16 to 64 bits. The range was 4.0 to 10.0 BRAMs for filters with kernel size $5\times 5$. On the other hand, the  $hls\_sobel$ utilized 9.0 BRAMs.  

DSP blocks are used in the multiplications required in the convolutions or for the polynomial approximations that compose the transcendental functions like division, exponentiation, square roots or logarithms. Note that the median filter did not use DSP blocks, as its architecture uses a sorting network that does not require multiplication. \textit{nlfilter} and \textit{fp\_sobel} used more DSP blocks, primarily because of the complex functions that require the polynomial approximations to compute the transcendental functions. It is essential to observe that the number of DSP blocks dropped for $conv_{5x5}$ when using $float64(53,10)$ and the filter consumed 206.20\% of the available LUTs in the Zybo Z7-20\footnote{The Zybo Z7-20 specification comprises the SoC XC7Z020-1CLG400C, which contains 53,200 LUTs, 106,400 flip-flops,  630 KB of Block RAM and 220 DSP blocks \cite{zybo-z7-20}.}, leading the implementation to fail. A similar trend also occurred in \textit{fp\_sobel}, which also failed the implementation with $float64(53,10)$ and consumed 135.08\% of the available LUTs. Overall, the floating-point Sobel used less hardware resource usage than its HLS version for custom floating-point widths of up to 24 bits, accentuating the idea that custom floating-point can outperform fixed-point hardware compactness in some applications.  

\section{Autogeneration of Custom Floating-point Cores using DSL}\label{sec:dsl}

The results discussed in section \ref{sec:results} showed that custom floating-point could outperform hardware compactness for some applications. However, at this point, the floating-point blocks that compose the spatial filters were complex structures that were difficult to implement.  

\begin{figure}[h]
    \centering 
    \includegraphics[width=0.7\linewidth]{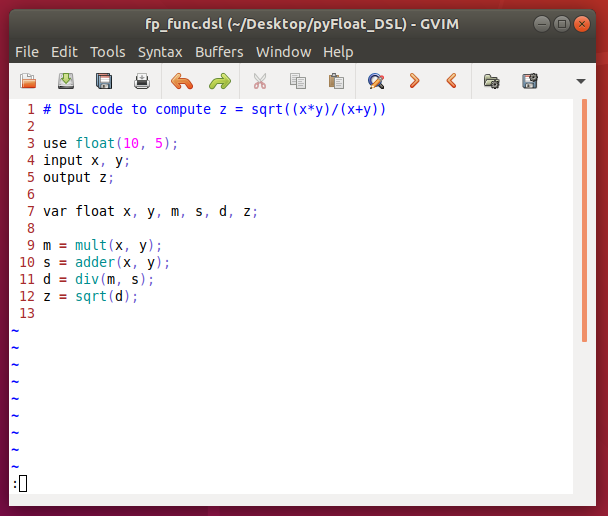}
    \caption{DSL code for the expression $z = \sqrt{\frac{(x\times y)}{(x+y)}}$ using $float16(10,5)$}
    \label{fig:fp_func_dsl}
\end{figure}

One way to optimize the tradeoff of rapid prototyping versus high-performance algorithms is using domain-specific languages (DSL). Employing a DSL with a syntax similar to Python or Matlab can produce complex algorithms with few lines of code. The compiler of the DSL parses lines of code with a high level of abstraction and translates abstract syntaxes into SystemVerilog instances of custom floating-point operations. The choice for custom floating-point arithmetic is driven by the fact that customizable floating-point can achieve a parameterizable dynamic range and precision that suits a particular application while saving hardware resources with the removal of unusable bits. Moreover, floating-point arithmetic helps designers produce faster architectures without specifying the bit width of intermediate variables in the datapath to avoid overflow.

A DSL compiler for custom floating-point was designed to accelerate the design of custom floating-point arithmetic focusing on image processing applications. The DSL compiler translates code from a domain-specific language into SystemVerilog instances of pipelined floating-point blocks. To exemplify the usage of the DSL, figure \ref{fig:fp_func_dsl} shows the DSL code implementation of the function $z=f(x,y)=\sqrt{\frac{(x\times y)}{(x+y)}}$.  The compiler receives the input file \texttt{fp\_func.dsl}, composed of 12 lines of code. The first line starts with a \texttt{\#} character, indicating that the line is a comment. Line 3 indicates that the arithmetic operations and variables are defined using a custom float with 10 bits for the mantissa and 5 bits for the exponent. Lines 4 and 5 declare the inputs and outputs of the function $z=f(x,y)$. Line 7 declares the variables, including inputs and outputs, used in the function with the type float. Lines 9 to 12 contain the operations to compute the function.

\begin{figure}[h]
    \centering 
    \includegraphics[width=0.73\linewidth]{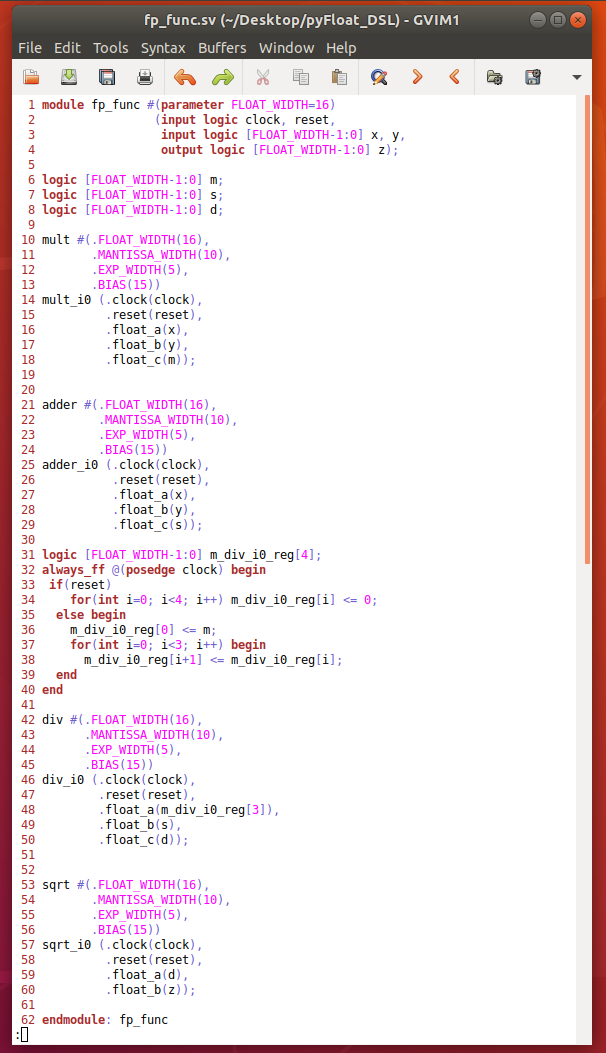}
    \caption{SystemVerilog code autogenerated from the DSL code in figure \ref{fig:fp_func_dsl}}
    \label{fig:fp_func_sv}
\end{figure}

From the DSL perspective, the code is sequential and untimed. For instance, lines 9 and 10 compute the product and sum of two inputs $x$ and $y$ and store those values in the variables $m$ and $s$, respectively. Therefore, $m$ and $s$ are the numerator and denominator of the quotient $d$, respectively, computed in line 11. Finally, line 12 computes the square root of the quotient $d$ and stores this value in $z$, which is the function's output. 

Although the explanation of the previous paragraph is elementary and looks unnecessary, the complexity of implementing the same function in hardware is complicated and tedious. The DSL compiler translates the code from figure \ref{fig:fp_func_dsl} into the code of figure \ref{fig:fp_func_sv}, which is a SystemVerilog pipelined implementation of the function $z=f(x,y)$ using $float16(10, 5)$. Unlike the DSL code, the generated code has a code density containing 62 lines. The code generation happens with the analysis of the DSL line by line. For instance, from line 3 in the DSL implementation, the compiler sets the values of \texttt{FLOAT\_WIDTH},\texttt{ MANTISSA\_WIDTH}, \texttt{EXP\_WIDTH} and \texttt{BIAS}, which are the parameters used in the floating-point blocks. 

In addition to the inputs and outputs declared in the DSL code, the compiler checks a database of each floating-point operation to insert additional inputs if necessary. For example, the \texttt{clock} and \texttt{reset} signals are required in the SystemVerilog instances, but they are omitted in the DSL code, which does not require the notion of time and schedule. The schedule of operations happens in the following way. For each line that contains an operation, the compiler checks the inputs and outputs of each operation. All the latencies of the signals are set to zero during the declaration of the variables. Output variables increase their latencies by the latency of each operation. For instance, after the multiplication of $x$ and $y$, $m$ has a latency of two cycles. Similarly, the latency of $s$ is increased by six cycles. Note that $x$ and $y$ remain with zero latencies, indicating that the addition and multiplication operations are performed in parallel, as opposed to the sequential operations from the DSL. However, the next operation takes $m$ and $s$ as inputs to compute the division. Because they were outputs of the two different operations with different latencies,  $m$ needs to be delayed by four clock cycles to match the latency of $s$. As already discussed in subsection \ref{subsec:generic_filter},  the number of cycles to delay $m$ is given by $\Delta(m, s) = max(\lambda(m), \lambda(s))-\lambda(s)=4$, where $\lambda(m)=2$ and $\lambda(s)=6$ are the latencies of $m$ and $s$, respectively. The compiler detects and checks the latency of each input signal. If the latencies of the inputs differ, then that latency is updated by $\Delta$ cycles to ensure the operation will compute correctly. In figure \ref{fig:fp_func_sv}, the signal \texttt{m\_div\_i0\_reg[3]} is inserted in the divider block at line 48.  \texttt{m\_div\_i0\_reg} is implemented in lines 31 to 40 and   \texttt{m\_div\_i0\_reg[i]} holds the value of $m$ delayed by $i+1$ clock cycles.  At this point, the result from the division is inputted into the square root block to compute the output result\footnote{The square root operation is computed using a degree-2, piecewise polynomial approximation and has a latency of five clock cycles. However, this notion of time is ignored in the DSL code.}. 

\begin{figure}[h]
    \centering 
    \includegraphics[width=0.75\linewidth]{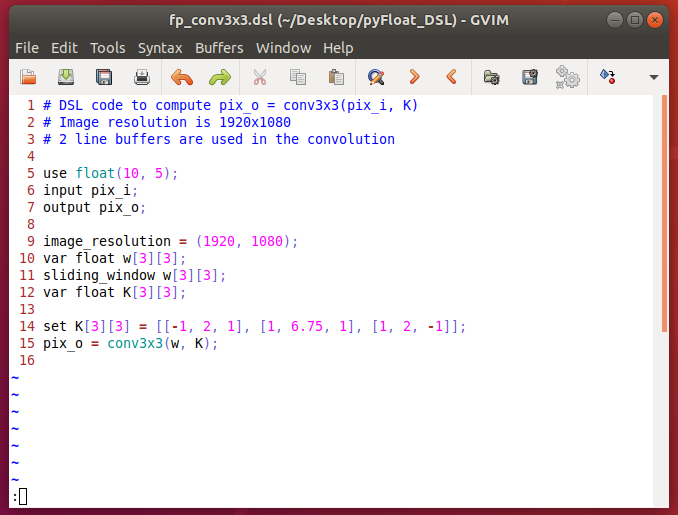}
    \caption{DSL code to implement a $3\times 3$ convolution}
    \label{fig:fp_conv3x3_dsl}
\end{figure}

Figure \ref{fig:fp_conv3x3_dsl} shows an example to implement a $3\times 3$ convolution with a $3\times 3$ kernel in $float16(10, 5)$. Line 9 in the DSL code sets the image resolution using the macro \texttt{image\_resolution}. The macro assigns two parameters \texttt{IMAGE\_HEIGHT = 1080} and \texttt{IMAGE\_WIDTH = 1920}.  Line 10 declares a floating-point array of \texttt{w[3][3]} dimensions $3\times 3$ that will be used in line 11, which in turn sets the variable $w$ to the output of a sliding window, that produces a window neighbourhood of dimensions $3\times 3$ every clock cycle. Moreover, the \texttt{sliding\_window} also sets another two parameters in the compiler: \texttt{WINDOW\_HEIGHT=3} and \texttt{WINDOW\_WIDTH=3}.  Thereafter, lines 12 and 14 declares and initialize a $3\times 3$ kernel. Finally, line 15 computes the convolution using the function $pix\_o = conv_{3\times 3}(pix\_i, K)$.\footnote{The function is defined in equation \ref{eq:lin_conv}.}

\begin{figure}[h]
    \centering 
    \includegraphics[width=0.73\linewidth]{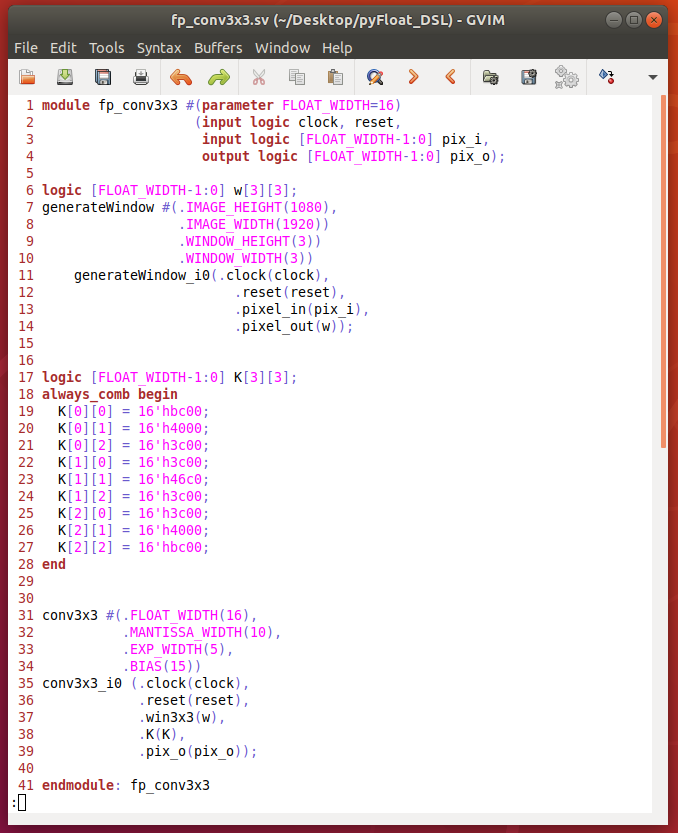}
    \caption{SystemVerilog code autogenerated from the DSL code in figure \ref{fig:fp_conv3x3_dsl}}
    \label{fig:fp_conv3x3_sv}
\end{figure}

The generated code for the $3\times 3$ convolution implemented in the DSL of figure \ref{fig:fp_conv3x3_dsl} is shown in figure  \ref{fig:fp_conv3x3_sv}. There, in lines 7 to 14, a module \texttt{generateWindow} is set with the parameters inputted from lines 9 to 11 in the DSL code. The module \texttt{generateWindow} generates a window neighbourhood \texttt{w} of $H\times W$ pixels using $H-1$ block RAMs. \texttt{w} and \texttt{K} are inputs of the $conv_{3\times 3}$. The Kernel \texttt{K}, initialized in the DSL code is transformed in the assignments inside the \texttt{always\_comb}, as can be seen in lines 18 to 28 in figure \ref{fig:fp_conv3x3_sv}. 

The floating-point representation is stored in hexadecimal format inside the \texttt{always\_comb}. The number transformation occurs as follows. For instance, the element \texttt{K[1][1]=6.75} can be written as \texttt{K[1][1]=6.75=$1.6875\times 2^2$}, which has the form of $x=(-1)^s\times 1.m\times 2^{e-BIAS}$. For a \texttt{float16} representation, the mantissa is stored as a 10-bit fixed-point value. Here, $m = \lfloor 0.6875\times 2^{10} \rfloor  = 704_{10} = 1011000000_2$. The exponent $exp=2+BIAS=17_{10}=10001_2$ and $s=0$. $x$ is formed as a concatenation of $x=(s, exp, m) = 0100011011000000_2$, which converted to hexadecimal gives the value $46c0$.

\begin{figure}[h]
    \centering 
    \includegraphics[width=0.73\linewidth]{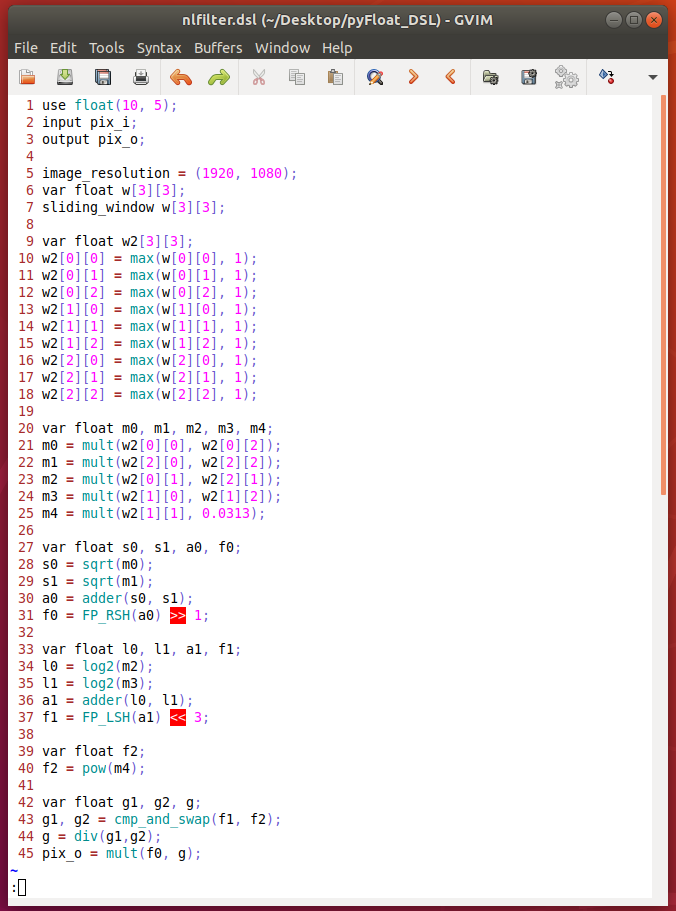}
    \caption{DSL code to implement the non-linear filter using generic function from equation \ref{eq:nl_funcs}}
    \label{fig:nlfilter_dsl}
\end{figure}

Figure \ref{fig:nlfilter_dsl} shows the DSL code to implement the non-linear filter\footnote{The SystemVerilog autogenerated contains 341 lines of code and is omitted due to formatting limitations.}. The filter receives the output of the sliding window \texttt{w[3][3]}. Line 9 declares a new array  \texttt{w2[3][3]}. The computation of $w2[i][j] = max(w[i][j], 1)$, for $0\leq i\leq 2$, $0\leq j\leq 2$ is done in lines 10 to 18. The maximum operation avoids the division by zero or the logarithm of zero.  Lines 28 and 36 contain the operations that sum the results of square roots and logarithms, computed in lines 28 to 29 and 34 to 35, respectively. The square roots and logarithms receive the results of the multiplications in lines 21 to 24. In line 31, the operation \texttt{f0 = FP\_RSH(a0) >> 1} is a floating-point shift to the right, which is equivalent to the expression  \texttt{f0 =2*a0}. Similarly, in line 37,  \texttt{f1 = FP\_LSH(a1) >> 1} equals  \texttt{f0 = 2*a1}. Using the floating-point shifters is efficient to compute multiplications and divisions by powers of two, as the floating-point shifters only increment or decrement the input by a constant. Still, in line 40, the function $2^{m4}$ is computed, where \texttt{m4} is the product of \texttt{w2[1][1]} by a constant value. Finally, in line 43, the operation \texttt{cmp\_and\_swap} will swap \texttt{f1} and \texttt{f2} if \texttt{f1>f2} and store the values in \texttt{g1} and \texttt{g2}. The division \texttt{g1/g2} is stored in \texttt{g} and multiplied by \texttt{f0}, to produce the output pixel. 

The DSL syntax allows the speedy implementation of the generic filter with the custom function defined in equation \ref{eq:nl_funcs}, enabling the rapid prototyping of spatial filters in FPGAs while reducing the error-prone tendency of manual hand-written RTL. Moreover, the custom floating-point arithmetic allows the exploration of precision vs hardware compactness. Therefore, the custom floating-point library, alongside the DSL compiler, will allow non-experts to quickly develop complex and efficient real-time image and video processing algorithms.

\section{Conclusion}\label{sec:conclusion}
Many algorithms require the operation of pixels in multiple lines. Therefore, a general window generator to compute sliding window operations was implemented in SystemVerilog. The window is of dimensions $H \times W$ and requires $H-1$ line buffers efficiently mapped to BRAMs in the FPGA. This structure enabled the implementation of different spatial filters, from linear convolutions to the generic filter with the custom function from equation \ref{eq:nl_funcs}.
Furthermore, the filters were implemented in different custom floating-point types with varying widths from 16 to 64 bits. The achieved throughput of the hardware implementation of the spatial filters was also observed when compared with software implementations. The non-linear filter with the custom function from equation \ref{eq:nl_funcs} evidenced the necessity of hardware implementations to meet real-time processing. Results showed that custom floating-point could outperform hardware compactness for some applications. However, at this point, the floating-point blocks that compose the spatial filters were complex structures that were difficult to implement. The DSL framework maintained the fast code generation, allowing the translation of a few lines of code containing untimed operations into complex pipelined architectures.

\section*{Acknowledgment}

The authors gratefully acknowledge the financial support of the Engineering and Physical Sciences Research Council (EPSRC) Centre for Doctoral Training in Embedded Intelligence under grant reference EP/L014998/1 and ARM Holdings for their support and input to this research.

\ifCLASSOPTIONcaptionsoff
  \newpage
\fi

\bibliography{references}

\begin{thebibliography}{10}
\providecommand{\url}[1]{#1}
\csname url@samestyle\endcsname
\providecommand{\newblock}{\relax}
\providecommand{\bibinfo}[2]{#2}
\providecommand{\BIBentrySTDinterwordspacing}{\spaceskip=0pt\relax}
\providecommand{\BIBentryALTinterwordstretchfactor}{4}
\providecommand{\BIBentryALTinterwordspacing}{\spaceskip=\fontdimen2\font plus
\BIBentryALTinterwordstretchfactor\fontdimen3\font minus
  \fontdimen4\font\relax}
\providecommand{\BIBforeignlanguage}[2]{{%
\expandafter\ifx\csname l@#1\endcsname\relax
\typeout{** WARNING: IEEEtran.bst: No hyphenation pattern has been}%
\typeout{** loaded for the language `#1'. Using the pattern for}%
\typeout{** the default language instead.}%
\else
\language=\csname l@#1\endcsname
\fi
#2}}
\providecommand{\BIBdecl}{\relax}
\BIBdecl

\bibitem{tsigkanos2020high}
A.~Tsigkanos, N.~Kranitis, D.~Theodoropoulos, and A.~Paschalis,
  ``High-performance cots fpga soc for parallel hyperspectral image compression
  with ccsds-123.0-b-1,'' \emph{IEEE Transactions on Very Large Scale
  Integration (VLSI) Systems}, vol.~28, no.~11, pp. 2397--2409, 2020.

\bibitem{ma2019performance}
Y.~Ma, Y.~Cao, S.~Vrudhula, and J.-S. Seo, ``Performance modeling for cnn
  inference accelerators on fpga,'' \emph{IEEE Transactions on Computer-Aided
  Design of Integrated Circuits and Systems}, vol.~39, no.~4, pp. 843--856,
  2019.

\bibitem{cong2018understanding}
J.~Cong, Z.~Fang, M.~Lo, H.~Wang, J.~Xu, and S.~Zhang, ``Understanding
  performance differences of fpgas and gpus,'' in \emph{2018 IEEE 26th Annual
  International Symposium on Field-Programmable Custom Computing Machines
  (FCCM)}.\hskip 1em plus 0.5em minus 0.4em\relax IEEE, 2018, pp. 93--96.

\bibitem{fingeroff2010high}
M.~Fingeroff, \emph{High-level synthesis: blue book}.\hskip 1em plus 0.5em
  minus 0.4em\relax Xlibris Corporation, 2010.

\bibitem{campos2017metodologia}
N.~C. d.~S. Campos, \emph{Uma metodologia de projeto e valida{\c{c}}{\~a}o de
  sistemas de detec{\c{c}}{\~a}o de faces.}\hskip 1em plus 0.5em minus
  0.4em\relax Universidade Federal de Campina Grande, 2017.

\bibitem{campos2017framework}
N.~C. Campos, H.~A. Monteiro, A.~V. Brito, A.~M. Lima, E.~U. Melcher, and M.~R.
  Morais, ``A framework for design and validation of face detection systems,''
  in \emph{2017 CHILEAN Conference on Electrical, Electronics Engineering,
  Information and Communication Technologies (CHILECON)}, 2017, pp. 1--7.

\bibitem{monteiro2017energy}
H.~A. Monteiro, N.~Campos, J.~P. Oliveira, A.~M.~N. Lima, A.~V. Brito, and
  E.~Melcher, ``Energy consumption measurement of a fpga full-hd video
  processing platform,'' in \emph{Proceedings of 7th Workshop on Circuits and
  Systems Design (WCAS’17), Fortaleza, Brazil}, 2017.

\bibitem{mohanty2012memory}
B.~K. Mohanty and P.~K. Meher, ``Memory-efficient high-speed convolution-based
  generic structure for multilevel 2-d dwt,'' \emph{IEEE Transactions on
  Circuits and Systems for Video Technology}, vol.~23, no.~2, pp. 353--363,
  2012.

\bibitem{holzer2012optimized}
M.~Holzer, F.~Schumacher, T.~Greiner, and W.~Rosenstiel, ``Optimized hardware
  architecture of a smart camera with novel cyclic image line storage
  structures for morphological raster scan image processing,'' in \emph{2012
  IEEE International Conference on Emerging Signal Processing
  Applications}.\hskip 1em plus 0.5em minus 0.4em\relax IEEE, 2012, pp. 83--86.

\bibitem{ioannou2020high}
L.~Ioannou, A.~Al-Dujaili, and S.~A. Fahmy, ``High throughput spatial
  convolution filters on fpgas,'' \emph{IEEE Transactions on Very Large Scale
  Integration (VLSI) Systems}, vol.~28, no.~6, pp. 1392--1402, 2020.

\bibitem{kent2020design}
R.~B. Kent and M.~S. Pattichis, ``Design, implementation, and analysis of
  high-speed single-stage n-sorters and n-filters,'' \emph{IEEE Access}, 2020.

\bibitem{saponara2005cost}
S.~Saponara, M.~Cassiano, S.~Marsi, R.~Coen, and L.~Fanucci, ``Cost-effective
  vlsi design of non linear image processing filters,'' in \emph{8th Euromicro
  Conference on Digital System Design (DSD'05)}.\hskip 1em plus 0.5em minus
  0.4em\relax IEEE, 2005, pp. 322--329.

\bibitem{fanucci2006asip}
L.~Fanucci, M.~Cassiano, S.~Saponara, D.~Kammler, E.~M. Witte, O.~Schliebusch,
  G.~Ascheid, R.~Leupers, and H.~Meyr, ``Asip design and synthesis for non
  linear filtering in image processing,'' in \emph{Proceedings of the Design
  Automation \& Test in Europe Conference}, vol.~2.\hskip 1em plus 0.5em minus
  0.4em\relax IEEE, 2006, pp. 1--6.

\bibitem{campos4}
\BIBentryALTinterwordspacing
{Campos, Nelson and Costa, Roberto and Costa, Elton and Junior, Gutemberg and
  Melcher, Elmar}, ``{A 4-MHz parameterized Logarithm-Square Root IP-Core},''
  12 2016. [Online]. Available:
  \url{https://www.design-reuse.com/articles/41709/a-4-mhz-parameterized-square-root-ip-core.html}
\BIBentrySTDinterwordspacing

\bibitem{campos2021fpga}
N.~Campos, S.~Chesnokov, E.~Edirisinghe, and A.~Lluis, ``Fpga implementation of
  custom floating-point logarithm and division,'' in \emph{International
  Symposium on Applied Reconfigurable Computing}.\hskip 1em plus 0.5em minus
  0.4em\relax Springer, 2021, pp. 295--304.

\bibitem{campos2022fpga}
N.~Campos, E.~Edirisinghe, S.~Fatima, S.~Chesnokov, and A.~Lluis, ``Fpga
  implementation of a custom floating-point library,'' in \emph{Proceedings of
  SAI Intelligent Systems Conference}.\hskip 1em plus 0.5em minus 0.4em\relax
  Springer, 2022, pp. 527--542.

\bibitem{De-Sousa-Campos2022}
\BIBentryALTinterwordspacing
N.~De-Sousa-Campos, ``{Framework for rapid hardware prototyping using custom
  floating-point arithmetic},'' 12 2022. [Online]. Available:
  \url{https://repository.lboro.ac.uk/articles/thesis/Framework_for_rapid_hardware_prototyping_using_custom_floating-point_arithmetic/21585504}
\BIBentrySTDinterwordspacing

\bibitem{schmuland2012cad}
T.~E. Schmuland, M.~M. Jamali, M.~B. Longbrake, and P.~E. Buxa, ``Cad tool
  autogeneration of vhdl fft for fpga/asic implementation,'' in \emph{10th IEEE
  International NEWCAS Conference}.\hskip 1em plus 0.5em minus 0.4em\relax
  IEEE, 2012, pp. 237--240.

\bibitem{serre2018dsl}
F.~Serre and M.~P{\"u}schel, ``A dsl-based fft hardware generator in scala,''
  in \emph{2018 28th International Conference on Field Programmable Logic and
  Applications (FPL)}.\hskip 1em plus 0.5em minus 0.4em\relax IEEE, 2018, pp.
  315--3157.

\bibitem{weinstein2007methodology}
R.~K. Weinstein, M.~S. Reid, and R.~H. Lee, ``Methodology and design flow for
  assisted neural-model implementations in fpgas,'' \emph{IEEE transactions on
  neural systems and rehabilitation engineering}, vol.~15, no.~1, pp. 83--93,
  2007.

\bibitem{mullapudi2015polymage}
R.~T. Mullapudi, V.~Vasista, and U.~Bondhugula, ``Polymage: Automatic
  optimization for image processing pipelines,'' \emph{ACM SIGARCH Computer
  Architecture News}, vol.~43, no.~1, pp. 429--443, 2015.

\bibitem{ragan2013halide}
J.~Ragan-Kelley, C.~Barnes, A.~Adams, S.~Paris, F.~Durand, and S.~Amarasinghe,
  ``Halide: a language and compiler for optimizing parallelism, locality, and
  recomputation in image processing pipelines,'' \emph{Acm Sigplan Notices},
  vol.~48, no.~6, pp. 519--530, 2013.

\bibitem{li2020heterohalide}
J.~Li, Y.~Chi, and J.~Cong, ``Heterohalide: From image processing dsl to
  efficient fpga acceleration,'' in \emph{Proceedings of the 2020 ACM/SIGDA
  International Symposium on Field-Programmable Gate Arrays}, 2020, pp. 51--57.

\bibitem{zhao2019automatic}
Y.~Zhao, X.~Gao, X.~Guo, J.~Liu, E.~Wang, R.~Mullins, P.~Y. Cheung,
  G.~Constantinides, and C.-Z. Xu, ``Automatic generation of multi-precision
  multi-arithmetic cnn accelerators for fpgas,'' in \emph{2019 International
  Conference on Field-Programmable Technology (ICFPT)}.\hskip 1em plus 0.5em
  minus 0.4em\relax IEEE, 2019, pp. 45--53.

\bibitem{reiche2014code}
O.~Reiche, M.~Schmid, F.~Hannig, R.~Membarth, and J.~Teich, ``Code generation
  from a domain-specific language for c-based hls of hardware accelerators,''
  in \emph{2014 international conference on hardware/software codesign and
  system synthesis (CODES+ ISSS)}.\hskip 1em plus 0.5em minus 0.4em\relax IEEE,
  2014, pp. 1--10.

\bibitem{membarth2015hipa}
R.~Membarth, O.~Reiche, F.~Hannig, J.~Teich, M.~K{\"o}rner, and W.~Eckert,
  ``Hipa cc: A domain-specific language and compiler for image processing,''
  \emph{IEEE Transactions on Parallel and Distributed Systems}, vol.~27, no.~1,
  pp. 210--224, 2015.

\bibitem{reiche2017generating}
O.~Reiche, M.~A. {\"O}zkan, R.~Membarth, J.~Teich, and F.~Hannig, ``Generating
  fpga-based image processing accelerators with hipacc,'' in \emph{2017
  IEEE/ACM International Conference on Computer-Aided Design (ICCAD)}.\hskip
  1em plus 0.5em minus 0.4em\relax IEEE, 2017, pp. 1026--1033.

\bibitem{bailey2011image}
D.~G. Bailey, ``Image border management for fpga based filters,'' in \emph{2011
  Sixth IEEE International Symposium on Electronic Design, Test and
  Application}.\hskip 1em plus 0.5em minus 0.4em\relax IEEE, 2011, pp.
  144--149.

\bibitem{mehta2011xilinx}
N.~Mehta, ``Xilinx 7 series fpgas embedded memory advantages,'' 2011.

\bibitem{bressert2012scipy}
E.~Bressert, ``Scipy and numpy: an overview for developers,'' 2012.

\bibitem{matlab_nlfilter}
MathWorks, ``{nlfilter: General sliding-neighborhood operations},''
  \url{https://www.mathworks.com/help/images/ref/nlfilter.html}, 2022, accessed
  on 18/06/2022.

\bibitem{zybo-z7-20}
{Digilent}, ``Zybo z7 reference manual,''
  \url{https://reference.digilentinc.com/reference/programmable-logic/zybo-z7/start},
  [Online; accessed 06-June-2019].

\end{thebibliography}
\bibliographystyle{IEEEtran}

\end{document}